\documentclass[12pt]{article}
\usepackage[T1]{fontenc}
\usepackage{a4wide}
\usepackage{amsmath}
\usepackage[dvips]{graphics}

\makeatletter

\newcommand{\LyX}{L\kern-.1667em\lower.25em\hbox{Y}\kern-.125emX\spacefactor1000}

\usepackage{amssymb}
\makeatother

\begin{document}

\renewcommand\theequation{\hbox{\normalsize\arabic{section}.\arabic{equation}}}

May 1998\hfill{}DFUB-98-10 

\vfill{}
{\centering \textbf{\LARGE Nonlinear Integral Equation and Finite Volume Spectrum
of Sine-Gordon Theory}\LARGE \par}
\vspace{1cm}

{\centering G. Feverati\footnote{
E-mail: feverati@bo.infn.it
}, F. Ravanini\footnote{
E-mail: ravanini@bo.infn.it
} and G. Tak{\' a}cs\footnote{
E-mail: takacs@bo.infn.it
}\par}
\vspace{0.5cm}

{\centering \emph{INFN Sezione di Bologna - Dipartimento di Fisica}\\
\emph{Via Irnerio 46}\\
\emph{40126 Bologna, Italy}\par}

\begin{abstract}
We examine the connection between the nonlinear integral equation (NLIE) derived
from light-cone lattice and sine-Gordon quantum field theory, considered as
a perturbed \( c=1 \) conformal field theory. After clarifying some delicate points
of the NLIE deduction from the lattice, we compare both analytic and numerical
predictions of the NLIE to previously known results in sine-Gordon theory. To
provide the basis for the numerical comparison we use data from Truncated Conformal
Space method. Together with results from analysis of infrared and ultraviolet
asymptotics, we find evidence that it is necessary to change the rule of quantization
proposed by Destri and de Vega to a new one which includes as a special case
that of Fioravanti et al. This way we find strong evidence for the validity
of the NLIE as a description of the finite size effects of sine-Gordon theory.\vfill{}

\end{abstract}
\clearpage

\section{Introduction}

\setcounter{equation}{0}

In the theory of \( 1+1 \) dimensional integrable models, the investigation of the spectrum
on a spacetime cylinder with finite spatial volume plays a very important role
in exploring finite size effects and provides a possibility to determine many
important physical characteristics. 

One of the first approaches to compute the finite volume spectrum was the the
Thermodynamic Bethe Ansatz (TBA) \cite{YY} which was used to calculate the vacuum (Casimir)
energy \cite{Zam-tba1}. The method was later extended to include ground states of charged
sectors \cite{Fendley}. More recently, using analytic properties of the TBA equations extended
for complex values of the volume parameter, an approach to get excited states
was proposed in \cite{tateo_dorey,blz,susy}. The method of \cite{tateo_dorey} to get excited states sheds light on the
analytic structure of the dependence of scaling functions on the spatial volume
and is the only method developed so far to deal with excited states in perturbations
of RCFTs. Its main drawbacks are that (1) it can be used only for systems for
which a TBA equation describing the vacuum in finite volume is known; (2) to
obtain the equation for a given excited state one has to do analytic continuation
for each case separately, and a major part of this continuation can only be
carried out numerically. Because of the requirement of the knowledge of the
vacuum TBA equation and the complications of the analytic continuation, this
method is limited at present to integrable perturbations of Virasoro minimal
models and other simple perturbed conformal field theories.

Another method developed to study the finite volume energy levels of \( 1+1 \) dimensional
field theories is the Truncated Conformal Space (TCS) approach \cite{yurzam}. Its big advantage
is that it makes possible to study the finite volume spectrum of any theory
for which a description in terms of perturbed conformal field theory is known
(integrability is not required). In practice, so far, it has been mainly used
to study perturbations of Virasoro minimal models. In this paper we use an extension
of the TCS method to perturbations of \( c=1 \) conformal field theories in order to
get a basis for comparison with the results obtained from the Non-Linear Integral
Equation (NLIE) method to be described below. The fundamental limitation of
this approach is that it only allows one to get numerical predictions for the
energy levels.

The method which is in the focus of this paper is that of the NLIE deduced from
integrable light cone lattice regularisation. The light cone lattice model was
proposed by Destri and de Vega in \cite{ddv-87} where a fermionic operator was constructed
on the lattice which was shown to satisfy the Thirring equations of motion in
the continuum limit. Therefore it was plausible to conjecture that the light
cone lattice has something to do with a regularisation of the sine-Gordon/massive
Thirring (sG/mTh) field theory. The states of the lattice model can be described
in a Bethe Ansatz framework which can be recast in the form of a nonlinear integral
equation using residue tricks. The equation describing the vacuum has been derived
in \cite{ddv-92,ddv-95} and it was shown that in the ultraviolet limit it reproduces the correct
value of the central charge \( c=1 \). We remark that similar methods were independently
introduced in Condensed Matter Physics by other authors \cite{klumper}. 

The first extension to describe the excited state spectrum was introduced by
Fioravanti et al. \cite{noi} where the spectrum of states containing only solitons (and
no antisolitons/breathers) has been described. An extension to generic excited
states of even topological charge was described by Destri and de Vega in \cite{ddv-97} which
however gave a different prescription from that of Fioravanti et al. when specialized
to the multi-soliton states examined in \cite{noi}. The contradiction was resolved in
\cite{noi3} where it was shown, using the ultraviolet conformal weights calculated from
the NLIE and a numerical comparison to TCS results, that the prescription given
by Destri and de Vega leads to states which are not contained in the Hilbert
space while the one proposed by Fioravanti et al. describes the spectrum of
pure soliton states correctly. 

The NLIE treated in this paper corresponds in algebraic terms to the nontwisted
affine Kac-Moody algebra \( A^{(1)}_{1}=\widehat{sl(2)} \). An extension of the approach to simply-laced algebras
of \( ADE \) type was first given in \cite{mariottini} for the case of the vacuum. More recently, in
\cite{zinn-justin} there appeared the extension to describe the spectrum of excited states.

In this paper we describe a generalization of our previous results in \cite{noi3} to generic
excited states with even value of the topological charge, by extending the framework
proposed by Destri and de Vega in \cite{ddv-97}. In doing so we show that their form of
the NLIE is not entirely correct and give a new derivation of the NLIE which
leads to the correct result and avoids certain subtle points which are present
in all of the derivations known up to now. We examine analytically both the
infrared and ultraviolet asymptotics of the scaling functions and show that
the quantisation prescription can be chosen in a way which is consistent with
the ultraviolet limit of the sG/mTh theory. We also give comparisons of the
scaling functions coming from the NLIE to the ones predicted by the TCS approach. 

The major motivation for studying the NLIE is that it gives a method of describing
all excited states in a single framework for all values of the (sine-Gordon)
coupling constant, in contrast to the TBA approach. We will return to a detailed
comparison of the two methods in the conclusions.

\section{\label{sG_CFT}The sine-Gordon/massive Thirring model as a perturbed CFT}

\subsection{c=1 CFT}

To fix our conventions and to define certain objects which are used later, we
give a brief summary of the \( c=1 \) free boson with a target space of a circle of
radius \( R \), which describes the UV limit of the theory. The Lagrangian of this
CFT is taken to be

\begin{equation}
\label{}
{\cal L}=\displaystyle\frac{1}{8\pi }\displaystyle\int ^{L}_{0}\partial _{\mu }\varphi (x,\, t)\partial ^{\mu }\varphi (x,\, t)dx\: ,\: x\in [0,L]\: ,
\end{equation}
where \( L \) is the spatial volume (i.e. the theory is defined on a cylindrical spacetime
with circumference \( L \)). In the sequel we will also use the complex Euclidean
coordinates \( z=e^{2\pi (t-ix)/L},\: \bar{z}=e^{2\pi (t+ix)/L} \). We classify the superselection sectors by the \( \widehat{U(1)}_{L}\times \widehat{U(1)}_{R} \) Kac-Moody symmetry
algebra, generated by the currents

\[
J(z)=i\partial _{z}\varphi \, ,\quad \bar{J}(\bar{z})=i\partial _{\bar{z}}\varphi \, .\]
The left/right moving energy-momentum tensor is given by
\[
T(z)=\frac{1}{8\pi }\partial _{z}\varphi \partial _{z}\varphi =\sum ^{\infty }_{k=-\infty }L_{k}z^{-k-2}\quad ,\quad \bar{T}(\bar{z})=\frac{1}{8\pi }\partial _{\bar{z}}\varphi \partial _{\bar{z}}\varphi =\sum ^{\infty }_{k=-\infty }\bar{L}_{k}\bar{z}^{-k-2}\]
The coefficients \( L_{n} \) and \( \bar{L}_{n} \) of the Laurent expansion of these fields generate two
mutually commuting Virasoro algebras. If we require the (quasi)periodic boundary
conditions

\[
\varphi (x+L,\, t)=\varphi (x,\, t)+2\pi mR\, ,\quad m\in \mathbb {Z}\, ,\]
then the sectors are labelled by a pair of integers \( (n,\, m) \), where \( \frac{n}{R} \) is the eigenvalue
of the total field momentum \( \pi _{0} \)

\[
\pi _{0}=\int\limits ^{L}_{0}\pi (x,\, t)dx\, ,\quad \pi (x,\, t)=\frac{1}{4\pi }\partial _{t}\varphi (x,\, t)\: ,\]
and \( m \) is the winding number, i.e. the eigenvalue of the topological charge \( Q \)
defined by

\[
Q=\frac{1}{2\pi R}\int\limits ^{L}_{0}\partial _{x}\varphi (x,\, t)dx\, .\]
In the sector with quantum numbers \( (n,\, m) \), the scalar field is expanded in modes
as follows: 
\[
\begin{array}{rl}
\displaystyle \varphi (x,t)= & \phi (z)+\bar{\phi }(\bar{z})\: ,\\
\phi (z)= & \displaystyle\frac{1}{2}\varphi _{0}-ip_{+}\log z+i\displaystyle\sum _{k\neq 0}\displaystyle\frac{1}{k}a_{k}z^{-k}\: ,\\
\bar{\phi }(\bar{z})= & \displaystyle\frac{1}{2}\varphi _{0}-ip_{-}\log \bar{z}+i\displaystyle\sum _{k\neq 0}\displaystyle\frac{1}{k}\bar{a}_{k}\bar{z}^{-k}\: ,
\end{array}\]
where the left and right moving field momenta \( p_{\pm } \) (which are in fact the two \( U(1) \)
Kac-Moody charges) are given by

\begin{equation}
\label{}
p_{\pm }=\displaystyle\frac{n}{R}\pm \displaystyle\frac{1}{2}mR\, .
\end{equation}
The Virasoro generators take the form

\[
L_{n}=\frac{1}{2}\sum ^{\infty }_{k=-\infty }:a_{n-k}a_{k}:\, ,\quad \bar{L}_{n}=\frac{1}{2}\sum ^{\infty }_{k=-\infty }:\bar{a}_{n-k}\bar{a}_{k}:\, ,\]
where the colons denote the usual normal ordering, according to which the oscillator
with the larger index is put to the right. 

The ground states of the different sectors \( (n,\, m) \) are created from the vacuum by
the (Kac-Moody) primary fields, which are vertex operators of the form

\begin{equation}
\label{}
V_{(n,\, m)}(z,\overline{z})=:\exp i(p_{+}\phi (z)+p_{-}\bar{\phi }(\overline{z})):\: .
\end{equation}
The left and right conformal weights of the field \( V_{(n,\, m)} \) (i.e. the eigenvalues of
\( L_{0} \) and \( \bar{L}_{0} \)) are given by the formulae

\begin{equation}
\label{UV_weights}
\Delta ^{\pm }=\displaystyle\frac{p^{2}_{\pm }}{2}.
\end{equation}
The Hilbert space of the theory is given by the direct sum of the Fock modules
built over the states

\begin{equation}
\label{}
\left| n,\, m\right\rangle =V_{(n,\, m)}(0,0)\left| vac\right\rangle \: ,
\end{equation}
with the help of the creation operators \( a_{-k}\: ,\: \bar{a}_{-k}\: k>0 \):

\[
{\cal H}=\bigoplus _{(n,\, m)}\{a_{-k_{1}}\ldots a_{-k_{p}}\bar{a}_{-l_{1}}\ldots \bar{a}_{-l_{q}}|n,\, m\rangle ,\, k_{1},\ldots \, k_{p},\, l_{1},\ldots \, l_{q}\in \mathbb {Z}_{+}\}\]
The Hamiltonian is expressed in terms of the Virasoro operators as

\begin{equation}
\label{}
H_{CFT}=\displaystyle\frac{2\pi }{L}\left( L_{0}+\overline{L}_{0}-\displaystyle\frac{c}{12}\right) \: ,
\end{equation}
where the central charge is \( c=1 \). The generator of spatial translations is given
by

\begin{equation}
\label{}
P=\displaystyle\frac{2\pi }{L}\left( L_{0}-\bar{L}_{0}\right) \; .
\end{equation}
The operator \( L_{0}-\bar{L}_{0} \) is the conformal spin which has eigenvalue \( nm \) on the primary field
\( V_{(n,\: m)} \).

One can also introduce twisted sectors using the operator \( {\cal T} \) of the spatial translation
with one period \( x\rightarrow x+L \). The primary fields \( V_{(n,\, m)} \) as defined above satisfy the periodicity
condition \( {\cal T}V_{(n,\, m)}=V_{(n,\, m)}. \) If we require the more general twisted boundary condition labelled
by a real parameter \( \nu  \) 
\[
{\cal T}V_{(n,\, m)}=\exp \left( i\nu Q\right) V_{(n,\, m)}\, ,\]
then we can generate superselection sectors for which \( n\in \mathbb {Z}+\frac{\nu }{2\pi } \). The only twisted boundary
condition occurring in this paper will be the one with \( \nu =\pi  \), which is necessary
to describe the fermions of the massive Thirring model. 

The classification of the (modular invariant) \( c=1 \) CFTs can be found in \cite{ginsparg} and is
not need in the sequel. We stress only that a particular \( c=1 \) CFT is specified
by giving the spectrum of the quantum numbers \( (n,m) \) such that the corresponding
set of vertex operators (and their descendants) forms a \emph{closed and local}
operator algebra. The locality requirement is equivalent to the fact that the
operator product expansions of any two such local operators is single valued
in the complex plane of \( z \). This condition, which is weaker than the modular
invariance of the CFT, is the adequate one since we consider the theory on a
spacetime cylinder and do not wish to define it on higher genus surfaces.

\subsection{\label{sG_mTh_relation}Sine-Gordon/massive Thirring field theory}

The Lagrangian of sine-Gordon theory is given by

\begin{equation}
\label{sG_Lagrangian}
{\cal L}_{sG}=\displaystyle\int \left( \displaystyle\frac{1}{2}\partial _{\nu }\Phi \partial ^{\nu }\Phi +\displaystyle\frac{\mu ^{2}}{\beta ^{2}}:\cos \left( \beta \Phi \right) :\right) dx\, ,
\end{equation}
where \( \Phi  \) denotes a real scalar field, while that of the massive Thirring theory
is of the following form:

\begin{equation}
\label{mTh_Lagrangian}
{\cal L}_{mTh}=\displaystyle\int \left( \bar{\Psi }(i\gamma _{\nu }\partial ^{\nu }+m_{0})\Psi -\displaystyle\frac{g}{2}\bar{\Psi }\gamma ^{\nu }\Psi \bar{\Psi }\gamma _{\nu }\Psi \right) dx\, ,
\end{equation}
describing a current-current selfinteraction of a Dirac fermion \( \Psi  \). It is known
that the two theories are deeply related provided their coupling constants satisfy

\[
\frac{\beta ^{2}}{4\pi }=\frac{1}{1+g/\pi }\, .\]
Both models can be considered as the perturbations of a  \( c=1 \) CFT by a potential
\( V \) :

\begin{equation}
\label{}
H_{sG/mTh}=H_{CFT}+V\quad ,\quad V=\lambda \displaystyle\int \left( V_{(1,0)}(z,\overline{z})+V_{(-1,0)}(z,\overline{z})\right) dx\: ,
\end{equation}
which is related to the bosonic Lagrangian (\ref{sG_Lagrangian}) by the following redefinitions
of the field and the parameters:

\begin{equation}
\label{}
\varphi =\sqrt{4\pi }\Phi \: ,\quad R=\displaystyle\frac{\sqrt{4\pi }}{\beta }\: ,\quad \lambda =\displaystyle\frac{\mu ^{2}}{\beta ^{2}}\: .
\end{equation}
For later convenience, we also define a new parameter \( p \) with

\begin{equation}
\label{}
p=\displaystyle\frac{\beta ^{2}}{8\pi -\beta ^{2}}\: .
\end{equation}
In terms of \( p \), \( p=1 \) is the free fermion point, and \( p=\displaystyle\frac{1}{k}\; ,\; k=1,2,\ldots  \) are the thresholds where a
new bound state (sG breather) appears. \( p<1 \) corresponds to the attractive and \( p>1 \)
to the repulsive regime. The potential term becomes marginal when \( \beta ^{2}=8\pi  \) which corresponds
to \( p=\infty  \). The perturbation conserves the topological charge \( Q \), which can be identified
with the usual topological charge of the sG theory and with the fermion number
of the mTh model.

The difference between the two theories is that they correspond to the perturbation
of two \emph{different local} c=1 CFTs (but by the same operator). The short
distance behaviour of the sG theory is described by the c=1 CFT with a local
operator algebra \( {\cal A}_{b} \) generated by the vertex operators 
\[
\{V_{(n,\, m)}:\, n,\, m\in \mathbb {Z}\}\, ,\]
all of which satisfy the periodic boundary condition and lead to a modular invariant
partition function. The primary fields of the UV limit of mTh theory are
\[
\{V_{(n,\, m)}:\, n\in \mathbb {Z},\, m\in 2\mathbb {Z}\: or\: n\in \mathbb {Z}+\frac{1}{2},\, m\in 2\mathbb {Z}+1\}\, ,\]
generating an operator algebra \( {\cal A}_{f} \), which also contains fields having antiperiodic
boundary conditions (\( \nu =\pi  \) and odd value of the topological charge), and corresponds
to a \( \Gamma _{2} \)-invariant partition function. In particular, the fields \( V_{(\pm 1/2,\pm 1)} \) have conformal
spin \( \frac{1}{2} \) and describe the UV limit of the one-fermion states. The interested reader
can find a detailed exposition in \cite{kl-me}. Note that the two algebras share a common
subspace with even values of the topological charge, generated by \( \{V_{(n,\, m)}:\, n\in \mathbb {Z},\, m\in 2\mathbb {Z}\} \), where the
massive theories described by the Lagrangians (\ref{sG_Lagrangian}) and (\ref{mTh_Lagrangian}) are identical. It
is exactly this subspace that is accessible with the NLIE technique and is discussed
in this paper.

\section{\label{NLIE_derivation}Derivation of the NLIE}

\setcounter{equation}{0}

Let us recall some basic facts about the NLIE. Consider a 1+1 Minkowski space-time
discretized along its light-cone directions, with (diagonal) lattice spacing
\( a \). The space-time has the geometry of a cylinder, of circumference \( L=Na \), where
\( N \) is the number of lattice sites in the spatial direction. On this lattice we
consider an inhomogeneous 6-vertex model, with an anisotropy parameter \( \gamma  \) and
periodic boundary conditions. The Boltzmann weights are realized in terms of
the quantum R-matrix, in fundamental representation for both the horizontal
and vertical spaces, of the \( U_{q}(\widehat{sl}(2)) \) affine algebra, \( q=-e^{-i\gamma } \). The inhomogeneities take the
values \( \Theta _{n}=(-1)^{n}\Theta /2 \) on the different spatial rows of the lattice, so that the unitary time
evolution operator can be written as
\[
U=e^{-iaH}=\tau ^{(2N)}\left( \frac{\Theta }{2}\Bigg |\{\Theta _{n}\}\right) \tau ^{(2N)}\left( -\frac{\Theta }{2}\Bigg |\{\Theta _{n}\}\right) ^{-1}\]
where \( \tau ^{(N)}(\vartheta |\{\Theta _{n}\}) \) denotes the usual 6-vertex transfer matrix on a square lattice of \( N \)
sites (see e.g. \cite{ddv-87} for details). 

The diagonalization of this evolution operator is achieved by Algebraic Bethe
Ansatz with the following set of Bethe equations
\begin{equation}
\label{bethe}
\left( \displaystyle\frac{\sinh \displaystyle\frac{\gamma }{\pi }\left[ \vartheta _{j}+\Theta +\displaystyle\frac{i\pi }{2}\right] \sinh \displaystyle\frac{\gamma }{\pi }\left[ \vartheta _{j}-\Theta +\displaystyle\frac{i\pi }{2}\right] }{\sinh \displaystyle\frac{\gamma }{\pi }\left[ \vartheta _{j}+\Theta -\displaystyle\frac{i\pi }{2}\right] \sinh \displaystyle\frac{\gamma }{\pi }\left[ \vartheta _{j}-\Theta -\displaystyle\frac{i\pi }{2}\right] }\right) ^{N}=-\prod _{k=1}^{M}\displaystyle\frac{\sinh \displaystyle\frac{\gamma }{\pi }\left[ \vartheta _{j}-\vartheta _{k}+i\pi \right] }{\sinh \displaystyle\frac{\gamma }{\pi }\left[ \vartheta _{j}-\vartheta _{k}-i\pi \right] }\: .
\end{equation}
\( M\leq N \) is the number of roots in the given Bethe configuration. Notice that the usual
requirement to have an even number of sites of a Bethe system (in order to be
able to define an antiferromagnetic vacuum) regards the number \( 2N \) appearing in
the transfer matrix \( \tau ^{(2N)} \), i.e. to the number of sites of the equivalent XXZ chain,
and \emph{not} to the number \( N \) itself, that can be chosen even or odd without
restriction. In general for a state with \( M \) roots the third component of the
spin of the chain is \( S=N-M \). The antiferromagnetic vacuum corresponds to the choice
of the maximal number of (all real) roots (a sort of Dirac sea), \emph{i.e.}
\( M=N \). The energy \( E \) and momentum \( P \) of a state can be obtained from the transfer
matrix eigenvalues by use of the formula

\begin{equation}
\label{autovalori}
e^{ia(E\pm P)/2}=(-1)^{M}\prod ^{M}_{j=1}\displaystyle\frac{\sinh \displaystyle\frac{\gamma }{\pi }\left[ \Theta \pm \vartheta _{j}+\displaystyle\frac{i\pi }{2}\right] }{\sinh \displaystyle\frac{\gamma }{\pi }\left[ \Theta \pm \vartheta _{j}-\displaystyle\frac{i\pi }{2}\right] }+O(1)\: .
\end{equation}
The Bethe equations (or rather their logarithms) can be reformulated as quantization
conditions for the \( \vartheta _{j} \)'s in terms of the so called \emph{counting function}, defined
as
\begin{equation}
\label{lattice_counting_function}
Z_{N}(\vartheta )=N[\phi _{1/2}(\vartheta +\Theta )+\phi _{1/2}(\vartheta -\Theta )]-\displaystyle\sum _{j=1}^{M}\phi _{1}(\vartheta -\vartheta _{k})\: ,
\end{equation}
where 
\[
\phi _{\nu }(\vartheta )=i\log \frac{\sinh \frac{\gamma }{\pi }(i\pi \nu +\vartheta )}{\sinh \frac{\gamma }{\pi }(i\pi \nu -\vartheta )}\: .\]
The branch of the logarithm in \( \phi _{\nu }(\vartheta ) \) is fixed by requiring analyticity and oddness
of the function in a strip around the real axis. Due to the periodicity of \( Z_{N}(\vartheta ) \)
evident from (\ref{lattice_counting_function}), the fundamental region to consider is 
\[
|\Im m\vartheta |\leq \frac{\pi ^{2}}{2\gamma }\: ,\]
which is wider than the analyticity strip. This is the reason for the introduction
of the so-called second determination of the counting function (cf. \cite{ddv-97}), to which
we come later. The roots of the Bethe equations must satisfy the condition
\begin{equation}
\label{quantum}
Z_{N}(\vartheta _{j})=2\pi I_{j}\qquad ,\qquad I_{j}\in \mathbb {Z}+\displaystyle\frac{1+\delta }{2}\: ,
\end{equation}
where \( \delta =M\bmod 2 \) so it takes the values \( 0 \) and \( 1 \). We return to a discussion of the value
of \( \delta  \) later. 

The physical renormalized continuum QFT on the cylinder is defined by sending
\( a\rightarrow 0 \) and \( N\rightarrow \infty  \) while keeping \( L=Na \) fixed. The renormalized mass scale \( {\cal M}=4a^{-1}e^{-\Theta } \) is generated by
sending the inhomogeneity \( \Theta  \), that can be interpreted as a momentum cutoff on
the lattice, to infinity together with \( N \), \emph{i.e.} 
\[
\Theta =\log \frac{4N}{l}\]
 (where \( l \) is the dimensionless scale \( l={\cal M}L \)). In \cite{ddv-87} it is shown that there exists
a certain fermionic operator on the lattice which satisfies equations of motion
that on the continuum become the Thirring ones. This supports the claim that
the continuum limit of this lattice theory is describing something related to
the sG/mTh models at coupling 
\[
\beta =\sqrt{8(\pi -\gamma )}\; \mathrm{or}\; p=\frac{\pi }{\gamma }-1\: .\]
It is interesting to see to what extent this approach can reproduce the spectrum
of sG/mTh states.

The number of roots of the system, which is of order \( N \) for the low level (non
thermodynamical) excitations we are interested in, becomes infinite in the continuum
limit. The Bethe equations also become infinite in number and are longer useful
to describe the states. Fortunately, the function \( Z_{N}(\vartheta ) \) has a finite limit. and
it satisfies a non-linear integral equation (NLIE in the following) already
in the lattice, which was constructed some years ago \cite{ddv-92,ddv-95} directly from the Bethe
equations. However, all the derivations reported so far, are plagued by one
or another imprecision. Therefore we reproduce here a new derivation that clarifies
some delicate points and also introduces naturally the corrections needed for
analyticity that were called \emph{special roots/holes} in ref.\cite{ddv-97}.

Before attacking this problem, we recall some facts about the solutions of Bethe
equations that are needed in the following. Real analyticity of the function
\( Z_{N}(\vartheta ) \) means that Bethe roots appear either as real roots or in complex conjugate
pairs (with the exception of the selfconjugate wide roots with \( \Im m\vartheta =\pi (p+1)/2 \) ). The vacuum
state, as said above, consists in \( M=N \) roots, all real. This number becomes infinite
on the continuum and can be seen as a sort of sea of real roots describing the
quantum vacuum of the theory. For excited states there can be some points \( h_{j} \)
on the real axis of \( \vartheta  \), where \( e^{iZ_{N}(h_{j})}=(-1)^{\delta +1} \), but that are \emph{not} roots of the original
Bethe equations. These points are known as \emph{holes}, and play the role of
particle excitations. Their number \( N_{H} \) is \( O(1) \) and not \( O(N) \) as for the real roots, so
it remains finite in the continuum limit. The same happens to the number of
complex roots. It is convenient to distinguish these latter into two classes:
the \emph{close} roots \( c_{j} \), \( j=1,...,M_{C} \) having 
\[
|\Im \mathrm{m}\, c_{j}|<\pi \min (1,p)\, ,\]
and the \emph{wide} roots \( w_{j} \), \( j=1,...,M_{W} \) with 
\[
\pi \min (1,p)<|\Im \mathrm{m}\, w_{j}|\leq \pi \frac{p+1}{2}\, .\]
Note that the set of wide roots include, in our conventions, also the \emph{selfconjugate}
roots appearing at 
\[
\Im m\, w_{j}=\pi \frac{p+1}{2}\, ,\]
whose complex conjugate is mapped in the root itself due to \( i\pi (p+1) \) periodicity of
the counting function. 

\( Z_{N}(\vartheta ) \) is real and finite on the real axis and so for every choice for \( I_{j} \)'s between
the minimum and maximum values of \( Z_{N}(\vartheta ) \) (\ref{quantum}) has at least on solution, which can
be either a real root or a hole.Among the real solutions \( x \) we distinguish the
\emph{normal} ones with \( Z'_{_{N}}(x)>0 \) and the \emph{special} ones with \( Z'_{_{N}}(x)<0 \). If \( y \) is a special
object (i.e. a special root or hole), then its quantum number \( I \) is ``degenerate'',
in the sense that there is at least another real solution \( x \) with the same quantum
number:

\[
Z_{N}(y)=2\pi I=Z_{N}(x)\, ,\quad y\neq x\: .\]
This is implied by the property that the \( Z_{N}(\vartheta ) \) function (for \( N \) not too small) is
``globally increasing''. From the same property it is obvious that the number
of solutions corresponding to a given value \( I \) is always odd.

\subsection{Non linear integral equation}

\noindent We write the counting function in the following way:
\begin{equation}
\label{count}
\displaystyle \begin{array}{rl}
\displaystyle Z_{N}(\vartheta ) & =N\left[ \phi _{1/2}(\vartheta +\Theta )+\phi _{1/2}(\vartheta -\Theta )\right] +\displaystyle\sum ^{N_{H}}_{k=1}\phi _{1}(\vartheta -h_{k})\\
 & -\displaystyle\sum ^{M_{C}+M_{W}}_{k=1}\phi _{1}(\vartheta -\xi _{k})-\displaystyle\sum ^{M_{R}+N_{H}}_{k=1}\phi _{1}(\vartheta -x_{k})\: ,
\end{array}
\end{equation}
where \( N_{H} \) is the total number of holes (normal and special), \( M_{R} \) the total number
of real roots (normal and special) and \( \xi _{k} \) denotes the set of the positions of
complex roots \( \{c_{j}\, ,w_{j}\} \). 

\noindent First we look for an equation for the derivative of the counting function
(\ref{count}). Let \( x \) be a real solution of the Bethe equation. We assume \( Z'_{_{N}}(x)\neq 0 \) and put, in
a complex neighbourhood of \( x \), \( \lambda =x+\epsilon  \). Then the expansion

\begin{equation}
\label{serie}
1+(-)^{\delta }e^{iZ_{N}(x+\epsilon )}\approx 1+(-)^{\delta }e^{iZ_{N}(x)+i\epsilon Z'_{_{N}}(x)}\approx -i\epsilon Z'_{_{N}}(x)
\end{equation}
has as obvious consequence the identity:

\begin{equation}
\label{identita}
\displaystyle\frac{1}{\lambda -x}=\displaystyle\frac{1}{\epsilon }=\displaystyle\frac{(-1)^{\delta }e^{iZ_{N}(x+\epsilon )}iZ'_{_{N}}(x+\epsilon )}{1+(-1)^{\delta }e^{iZ_{N}(x+\epsilon )}}+...={\cal R}_{\delta }(x+\epsilon )+...
\end{equation}
 (the dots are regular terms in \( \lambda -x \)). In the following we use the shorthand notation
\[
{\cal R}_{\delta }(\vartheta )=\frac{(-1)^{\delta }e^{iZ_{N}(\vartheta )}iZ'_{_{N}}(\vartheta )}{1+(-1)^{\delta }e^{iZ_{N}(\vartheta )}}\: .\]
One could be tempted to say (as in \cite{ddv-97}) that \( {\cal R}_{\delta }(\vartheta ) \) is just the derivative of 
\[
\log \left( 1+(-1)^{\delta }e^{iZ_{N}(\vartheta )}\right) \, \, ,\]
but this is true only while the logarithm remains in its fundamental domain
which is in general not true in the lower half plane. This is why we prefer
to use the function \( {\cal R}_{\delta }(\vartheta ) \) without expressing it as a logarithmic derivative that
could lead to abuse, for example, of integrations by parts.\footnote{
This is exactly the origin of unconsistencies in the derivation of \cite{ddv-97}. One should
either fix the logarithm to be in the fundamental domain and pay attention to
avoid branch cuts all along the integration contour (as we do in what follows)
or introduce a ``contour logarithm'' defined by integrating the function \( {\cal R}_{\delta } \) along
a suitable path taken in the complex plane, but then one needs to pay attention
to logarithmic identities and the use of integration by parts. 
}

The basic trick of the NLIE derivation \cite{ddv-95,noi,ddv-97} is to use the Cauchy theorem and identity
(\ref{identita}), to express an analytic function \( f(x) \) as

\begin{equation}
\label{cauchy}
f(x)=\oint _{\Gamma _{x}}\displaystyle\frac{d\lambda }{2\pi i}\displaystyle\frac{f(\lambda )}{\lambda -x}=\oint _{\Gamma _{x}}\displaystyle\frac{d\lambda }{2\pi i}f(\lambda ){\cal R}_{\delta }(\lambda )\, ,
\end{equation}
where \( \Gamma _{x} \) is a curve encircling \( x \) anti-clockwise, while avoiding other singularities
of the denominator i.e. other Bethe solutions (real or complex). This is always
possible because the solutions are finite in number as long as we are on the
lattice with \( N<\infty  \). Applying the same argument applies to each one of the real roots
of (\ref{bethe}) the last sum in the derivative of (\ref{count}) can be expressed as:
\[
\sum ^{M_{R}+N_{H}}_{k=1}\phi '_{_{\gamma }}(\vartheta -x_{k})=\oint _{\Gamma }\frac{d\lambda }{2\pi i}\phi '_{_{\gamma }}(\vartheta -\lambda ){\cal R}_{\delta }(\lambda )\: .\]
The sum on the contours was modified to a single curve \( \Gamma  \), depicted in figure
1, encircling all the real roots \( \{x_{k}\} \), and avoiding the complex Bethe solutions.

\begin{figure}
{\centering \resizebox*{0.8\textwidth}{0.3\textheight}{\rotatebox{270}{\includegraphics{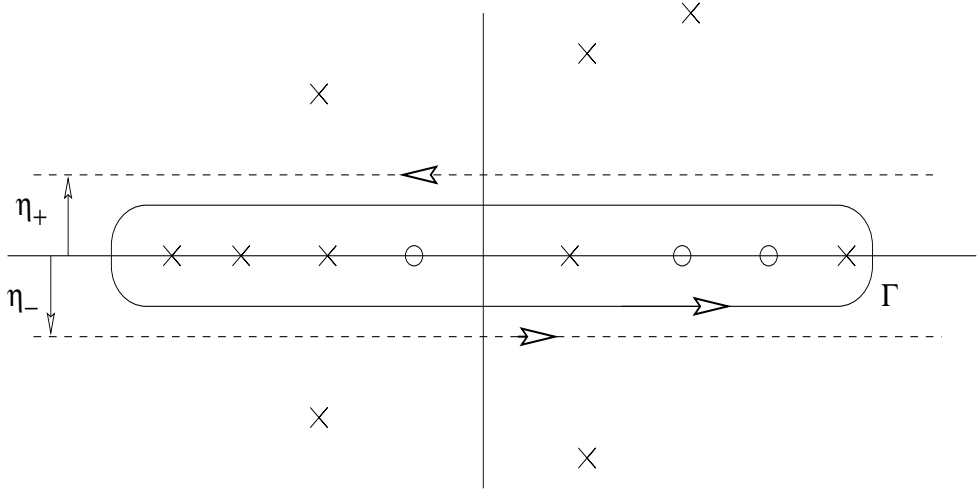}}} \par}

\caption{\small \label{contour} Contour for the integration. The crosses are roots while the circles are
holes.}
\end{figure}We take \( \Gamma  \) inside the strip \( -\eta _{-}\leq \Im m\vartheta \leq \eta _{+} \)(figure \ref{contour}), where \( \eta _{\pm } \) are positive real numbers satisfying

\[
0<\eta _{+},\eta _{-}\leq \frac{\pi }{2}\min \{1,\: p,|\Im m\: \xi _{k}|\, ,\, k=1,...,M_{C}+M_{W}\}\: .\]
Without loss of generality we can assume that \( \eta _{+}=\eta _{-}=\eta  \), and deform \( \Gamma  \) to the boundaries
of the whole strip. The regions at \( \pm \infty  \) do no contribute because \( \phi ' \) vanishes, and
so the integral can be performed on the lines \( \lambda =\rho \pm i\eta  \), where \( \rho  \) is real:
\[
\begin{array}{rl}
\displaystyle \oint _{\Gamma }\displaystyle\frac{d\lambda }{2\pi i}\phi '_{_{\gamma }}(\vartheta -\lambda ){\cal R}_{\delta }(\lambda ) & =\displaystyle\int\limits ^{+\infty }_{-\infty }\displaystyle\frac{d\rho }{2\pi i}\phi '_{_{\gamma }}(\vartheta -\rho +i\eta ){\cal R}_{\delta }(\rho -i\eta )\\
 & -\displaystyle\int\limits ^{+\infty }_{-\infty }\displaystyle\frac{d\rho }{2\pi i}\phi '_{_{\gamma }}(\vartheta -\rho -i\eta ){\cal R}_{\delta }(\rho +i\eta )\: .
\end{array}\]
After some algebraic manipulations similar to those in \cite{ddv-97} one obtains a NLIE
for the \( Z'_{_{N}}(\vartheta ) \) function
\begin{equation}
\label{zetap}
\begin{array}{rl}
\displaystyle Z'_{_{N}}(\vartheta ) & =N\left[ \displaystyle\frac{1}{\cosh (\vartheta +\Theta )}+\displaystyle\frac{1}{\cosh (\vartheta -\Theta )}\right] +\displaystyle\sum ^{N_{H}}_{k=1}2\pi G(\vartheta -h_{k})\\
 & -\displaystyle\sum ^{M_{C}}_{k=1}2\pi G(\vartheta -c_{k})-\displaystyle\sum ^{M_{W}}_{k=1}2\pi G(\vartheta -w_{k})_{II}\\
 & +2\Im m\displaystyle\int\limits ^{+\infty }_{-\infty }d\rho G(\vartheta -\rho -i\eta ){\cal R}_{\delta }(\rho +i\eta )\: ,
\end{array}
\end{equation}
where 
\[
G(\vartheta )=\frac{1}{2\pi }\int\limits ^{+\infty }_{-\infty }dk\, e^{ik\vartheta }\frac{\sinh \frac{\pi (p-1)k}{2}}{2\sinh \frac{\pi pk}{2}\, \cosh \frac{\pi k}{2}}\]
and \( G_{II} \) denotes the so-called \emph{second determination} of the function \( G \). Using
the prescription given in \cite{ddv-97}, the second determination of any function \( f(\vartheta ) \) is given
by 
\begin{equation}
\label{2nd_determination}
f(\vartheta )_{II}=\left\{ \begin{array}{ll}
f(\vartheta )+f\left( \vartheta -i\pi \mathrm{sign}\left( \Im m\vartheta \right) \right) \: , & p>1\: ,\\
f(\vartheta )-f\left( \vartheta -ip\pi \mathrm{sign}\left( \Im m\vartheta \right) \right) \: , & p<1\: .
\end{array}\right. \: .
\end{equation}

Equation (\ref{zetap}) holds for both positive and negative values of \( Z'_{_{N}} \). We already noted
that the function \( {\cal R}_{\delta } \) is the derivative of 
\[
\log \left[ 1+(-1)^{\delta }e^{iZ_{N}(x+i\eta )}\right] \, \, ,\]
but some care is required because of the multivaluedness of the logarithmic
function. The argument of the logarithm, calculated at a real solution of (\ref{bethe}),
is (see (\ref{serie}))
\begin{equation}
\label{sviluppo}
1+(-1)^{\delta }e^{iZ_{N}(x+\epsilon +i\eta )}\approx (-i\epsilon +\eta )Z'_{_{N}}(x)\: .
\end{equation}
It intersects the negative real line which is the cut of the logarithm, if \( Z'_{_{N}}(x)<0 \),
which corresponds to the case of special roots and holes. If there are no special
objects, the argument of the logarithm always has a positive real part, without
crossing the logarithmic cut, thus we can write the integral as:

\[
2\Im m\int\limits ^{+\infty }_{-\infty }dxG(\lambda -x-i\eta )\frac{d}{dx}\log _{FD}\left[ 1+(-1)^{\delta }e^{iZ_{N}(x+i\eta )}\right] \: ,\]
where \( \log _{FD} \) denotes the fundamental determination of the logarithm and the integral
runs along the real axis. 

\begin{figure}
{\centering \resizebox*{0.8\textwidth}{!}{\rotatebox{270}{\includegraphics{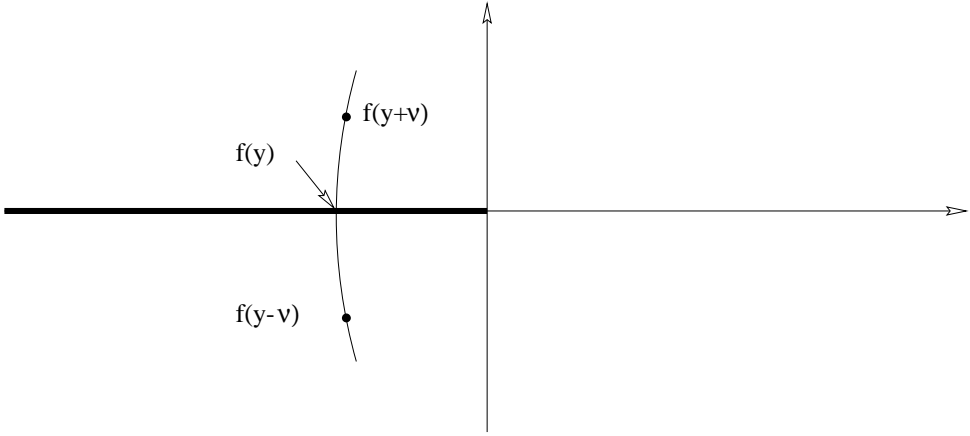}}} \par}

\caption{\small \label{special}Values of the function \protect\( f(x)\protect \); \protect\( y\protect \) is a special object and \protect\( \nu \protect \) is a positive real number.}
\end{figure}

For real negative \( w \), \( \log (w+i0^{_{+}})-\log (w-i0^{_{-}})=2\pi i \) for any choice of the logarithmic branch. This is important
for special objects \( y_{k} \) as evident from (\ref{sviluppo}). Let us denote \( f(x)=1+(-1)^{\delta }e^{iZ_{N}(x+i\eta )} \) (with \( x \) real) and
consider the simplest case where there is only one special object \( y \) (\( Z'_{_{N}}(y)<0 \), \( Z_{N}(y)=2\pi I \) and
\( y\in \mathbb {R} \)). Observe that \( \log _{FD}f(x) \) is discontinuous in \( y \), as shown in figure \ref{special}. However, the
factor \( {\cal R}_{\delta }(x+i\eta ) \) in (\ref{zetap}) is continuous everywhere and so it must be the derivative of
a continuous function, namely of \( \log _{FD}f(x)-2\pi i\theta (x-y) \) (plus an arbitrary constant).

In the presence of \( N_{S} \) special objects we obtain
\[
{\cal R}_{\delta }(x+i\eta )=\frac{d}{dx}\log _{FD}\left( 1+(-1)^{\delta }e^{iZ_{N}(x+i\eta )}\right) -\sum _{k=1}^{N_{S}}2\pi i\delta (x-y_{k})\quad .\]
In the following, we omit the label \( FD \) and \( \log  \) will always denote the logarithm
in its fundamental branch. It is very important to note that the special objects
are not independent degrees of freedom unlike the holes and complex solutions
that are fixed ``a priori'': the special objects appear when the derivative of
\( Z_{N} \) at a root or at a hole becomes negative. 

Now we can come back to (\ref{zetap}) and write it with the explicit inclusion of special
objects
\begin{equation}
\label{zetaprimo}
\begin{array}{rl}
\displaystyle Z'_{_{N}}(\vartheta ) & =N\left[ \displaystyle\frac{1}{\cosh (\vartheta +\Theta )}+\displaystyle\frac{1}{\cosh (\vartheta -\Theta )}\right] +\displaystyle\sum ^{N_{H}}_{k=1}2\pi G(\vartheta -h_{k})+\\
 & -2\displaystyle\sum ^{N_{S}}_{k=1}2\pi G(\vartheta -y_{k})-\displaystyle\sum ^{M_{C}}_{k=1}2\pi G(\vartheta -c_{k})-\displaystyle\sum ^{M_{W}}_{k=1}2\pi G(\vartheta -w_{k})_{II}+\\
 & +2\Im m\displaystyle\int\limits ^{\infty }_{-\infty }d\rho G(\vartheta -\rho -i\eta )\displaystyle\frac{d}{d\rho }\log \left( 1+(-1)^{\delta }e^{iZ_{N}(\rho +i\eta )}\right) \: .
\end{array}
\end{equation}
The odd primitive of \( G(\vartheta ) \)

\begin{equation}
\label{chi}
\chi (\vartheta )=2\pi \displaystyle\int\limits _{0}^{\vartheta }dxG(x)=\displaystyle\int\limits ^{\infty }_{-\infty }\displaystyle\frac{dk}{ik}e^{ik\vartheta }\displaystyle\frac{\sinh \displaystyle\frac{\pi (p-1)k}{2}}{2\sinh \displaystyle\frac{\pi pk}{2}\cosh \displaystyle\frac{\pi k}{2}}
\end{equation}
has the curious property that it equals the logarithm of the soliton-soliton
sG scattering amplitude \( S_{++}^{++}(\vartheta ) \) as remarked in \cite{ddv-97}. This is much more than a coincidence
and as we shall see in section \ref{IR_limit} that it is central in reproducing the sine-Gordon
scattering amplitudes.

Integrating in \( \vartheta  \), we obtain the \textit{fundamental non linear integral equation}
for the counting function:

\begin{equation}
\label{nlie}
\begin{array}{rl}
\displaystyle Z_{N}(\vartheta ) & =2N\arctan \displaystyle\frac{\sinh \vartheta }{\cosh \Theta }+\displaystyle\sum ^{N_{H}}_{k=1}\chi (\vartheta -h_{k})-2\displaystyle\sum ^{N_{S}}_{k=1}\chi (\vartheta -y_{k})-\\
 & -\displaystyle\sum ^{M_{C}}_{k=1}\chi (\vartheta -c_{k})-\displaystyle\sum ^{M_{W}}_{k=1}\chi (\vartheta -w_{k})_{II}+\\
 & +2\Im m\displaystyle\int\limits ^{\infty }_{-\infty }d\rho G(\vartheta -\rho -i\eta )\log \left( 1+(-1)^{\delta }e^{iZ_{N}(\rho +i\eta )}\right) +C\: .
\end{array}
\end{equation}
The integration of the convolution term is possible because \( G \) vanishes exponentially
at infinity. \( C \) is the integration constant and must be determined by a comparison
with the asymptotic values of \( Z_{N} \) calculated by the definition (\ref{count}).

\subsection{\label{continuum_limit}Continuum limit}

To perform the continuum limit we send \( N,\, \Theta \rightarrow \, \infty  \) in the following way:

\[
\Theta \approx \log \frac{4N}{{\cal M}L}\: ,\]
as explained in \cite{ddv-92}. To obtain the energy and momentum eigenvalues in the continuum
limit requires a \emph{continuum counting function} that we define as

\[
Z(\vartheta )=\lim _{N\, \rightarrow \, \infty }Z_{N}(\vartheta )\: .\]
We obtain by performing the limit on (\ref{nlie})
\begin{equation}
\label{nlie-cont}
\displaystyle Z(\vartheta )=l\sinh \vartheta +g(\vartheta |\vartheta _{j})+2\Im m\displaystyle\int\limits ^{\infty }_{-\infty }dxG(\vartheta -x-i\eta )\log \left( 1+(-1)^{\delta }e^{iZ(x+i\eta )}\right) +\tilde{C}\: ,
\end{equation}
where \( l={\cal M}L \) is the dimensionless volume parameter. The function \( g(\vartheta |\vartheta _{j}) \) is the so-called
\emph{source term}, composed of the contributions from the holes, special objects
and complex roots which we call \emph{sources} and denote their positions by
the general symbol \( \{\vartheta _{j}\}=\{h_{k}\, ,\, y_{k}\, ,\, c_{k}\, ,\, w_{k}\} \)and takes the form
\[
\displaystyle g(\vartheta |\vartheta _{j})=\sum ^{N_{H}}_{k=1}\chi (\vartheta -h_{k})-2\sum ^{N_{S}}_{k=1}\chi (\vartheta -y_{k})-\sum ^{M_{C}}_{k=1}\chi (\vartheta -c_{k})-\sum ^{M_{W}}_{k=1}\chi (\vartheta -w_{k})_{II}\: .\]

The integration constant gets redefined by the asymptotic value of source terms
which disappear at the two infinities in the continuum limit. The value of the
constant \( \tilde{C} \) can be obtained using the fact the function \( Z \) as defined by (\ref{count}) is
asymptotically odd modulo \( 2\pi  \). The integral term goes to \( 0 \) if \( \vartheta \rightarrow \pm \infty  \), while the source
terms are odd and so we derive that \( \tilde{C}=0 \) modulo \( 2\pi  \) and so we are free to choose
\( \tilde{C}=0 \). The only problematic point arises when there are selfconjugate roots, for
which we will work around the problem by slightly changing the definition of
\( \chi _{II} \) with respect to the one given in \cite{ddv-97}. From the formula (\ref{2nd_determination}) in the repulsive
regime we obtain that the source term for a selfconjugate root (at position
\( w \) with \( \Im m\, w=\pi (p+1)/2 \)) is
\[
-i\log \left( -\frac{\cosh \frac{1}{2p}\left( 2(\vartheta -\Re e\, w)+i\pi \right) }{\cosh \frac{1}{2p}\left( 2(\vartheta -\Re e\, w)-i\pi \right) }\right) \: .\]
In choosing the logarithmic branches, we require that this function be continuous,
but then it cannot be odd. Therefore we modify the source term for this case
and \emph{define} it to be
\[
-i\log \frac{\cosh \frac{1}{2p}\left( 2(\vartheta -\Re e\, w)+i\pi \right) }{\cosh \frac{1}{2p}\left( 2(\vartheta -\Re e\, w)-i\pi \right) }\: .\]
This is equivalent to shifting \( Z \) by an odd multiple of \( \pi  \). The nice feature is
that we can then set \( \tilde{C}=0 \) for all source configurations. This redefinition will
come up again in section \ref{UV_limit}, where it will allow us to write the result for the
attractive and the repulsive regime in a unified form. The notation \( \chi _{II} \) will be
used without change but with the above redefinition understood where necessary.

The equation reported in \cite{ddv-97} has the form
\begin{equation}
\label{nlie-wrong}
\begin{array}{rl}
\displaystyle Z(\vartheta )= & l\sinh \vartheta +g(\vartheta |\vartheta _{j})-i\displaystyle\int\limits ^{\infty }_{-\infty }dxG(\vartheta -x-i\eta )\log \left( 1+(-1)^{\delta }e^{iZ(x+i\eta )}\right) +\\
 & +i\displaystyle\int\limits ^{\infty }_{-\infty }dxG(\vartheta -x+i\eta )\log \left( (-1)^{\delta }+e^{-iZ(x-i\eta )}\right) \: .
\end{array}
\end{equation}
This equation differs from ours by the position of the term \( (-1)^{\delta } \) in the second
logarithm. This difference, however small it appears, has profound implications.
First of all, contrary to the claims made by the authors of \cite{ddv-97}, for \( \delta =1 \), the second
term of (\ref{nlie-wrong}) does have logarithmic branch ambiguities even in the absence of
special objects; it is enough to examine its behaviour substituting \( Z(\vartheta )=l\sinh \vartheta  \) which
is the limit of the counting function for very large \( l \). 

The second point is that with our form of the equation the convolution term
decays exponentially fast with \( l \). This is important because in the large \( l \) limit
the source terms allow us to reconstruct an asymptotic Bethe Ansatz based on
the S-matrix of the continuum theory (see section \ref{IR_limit} for details). With the equation
(\ref{nlie-wrong}) the contribution coming from the convolution term decays as a negative power
of \( l \) if \( \delta =1 \). However, from general principles we know that the corrections to
the asymptotic Bethe Ansatz come from correlation effects which should decay
exponentially with the distance in a massive field theory.

Finally, taking the \( \eta \rightarrow 0 \) limit one can reconstruct the original Bethe equations
from (\ref{nlie}). For that reason one has to examine the function 
\[
{\cal Q}_{N}(\vartheta )=\mathop {\lim }_{\eta \rightarrow 0}\frac{1}{i}\log \frac{1+(-1)^{\delta }e^{iZ_{N}(\vartheta +i\eta )}}{1+(-1)^{\delta }e^{-iZ_{N}(\vartheta -i\eta )}}\: .\]
With our formulas we obtain
\begin{equation}
\label{Q_Z_relation}
{\cal Q}_{N}(\vartheta )=\left( Z_{N}(\vartheta )+\pi \delta \right) \bmod 2\pi \: ,
\end{equation}
which has jumps of \( 2\pi  \) exactly where the value of \( \vartheta  \) coincides with the position
of a real root or hole. By a partial integration one obtains \( \delta  \)-function terms
and gets back to the original definition of the counting function (\ref{count}). 

On the contrary, using the formula (\ref{nlie-wrong}) one arrives at 
\[
{\cal Q}_{N}(\vartheta )=\mathop {\lim }_{\eta \rightarrow 0}\frac{1}{i}\log \frac{1+(-1)^{\delta }e^{iZ_{N}(\vartheta +i\eta )}}{(-1)^{\delta }+e^{-iZ_{N}(\vartheta -i\eta )}}=Z_{N}(\vartheta )\bmod 2\pi \: ,\]
which is wrong for the case \( \delta =1 \) because the jumps are not at the positions of
real roots/holes. The authors of \cite{ddv-97} use the formula (\ref{Q_Z_relation}) which is the correct
one (cf. eqn. (4.14) in their paper), but it does not follow from their form
of the NLIE.

The continuum limits for energy and momentum are the similar to the ones exposed
in \cite{ddv-97}. The expression for energy, subtracting the bulk contribution is:
\begin{equation}
\label{energy}
\begin{array}{rl}
\displaystyle E-E_{bulk} & ={\cal M}\displaystyle\sum ^{N_{H}}_{j=1}\cosh h_{j}-2{\cal M}\displaystyle\sum ^{N_{S}}_{j=1}\cosh y_{j}\\
 & -{\cal M}\displaystyle\sum ^{M_{C}}_{j=1}\cosh c_{j}+{\cal M}\displaystyle\sum _{j=1}^{M_{W}}(\cosh w_{j})_{II}\\
 & -{\cal M}\displaystyle\int\limits ^{\infty }_{-\infty }\displaystyle\frac{dx}{2\pi }2\Im m\left[ \sinh (x+i\eta )\log (1+(-1)^{\delta }e^{iZ(x+i\eta )})\right] \: ,
\end{array}
\end{equation}
while for the momentum we get
\begin{equation}
\label{momentum}
\begin{array}{rl}
\displaystyle P & ={\cal M}\displaystyle\sum ^{N_{H}}_{j=1}\sinh h_{j}-2{\cal M}\displaystyle\sum ^{N_{S}}_{j=1}\sinh y_{j}\\
 & -{\cal M}\displaystyle\sum ^{M_{C}}_{j=1}\sinh c_{j}+{\cal M}\displaystyle\sum _{j=1}^{M_{W}}(\sinh w_{j})_{II}\\
 & -{\cal M}\displaystyle\int\limits ^{\infty }_{-\infty }\displaystyle\frac{dx}{2\pi }2\Im m\left[ \cosh (x+i\eta )\log (1+(-1)^{\delta }e^{iZ(x+i\eta )})\right] \: ,
\end{array}
\end{equation}
where the values of \( h_{j},\, c_{j},\, y_{j} \) and \( w_{j} \) are fixed by the quantization conditions (\ref{quantum}). 

The input required to determine a state is given by the number of holes, close
and wide pairs and their quantum numbers \( I_{j} \). It is not necessary to specify the
special sources: since \( Z'(\vartheta ) \) is dominated by the term \( l\cosh \vartheta  \) for large \( l \), it is always
positive and so the special sources disappear. As a consequence, to specify
a state one can give only the normal sources at far infrared. The need for the
special sources can be detected by gradually decreasing the value of \( l \) and looking
at the sign of the derivative of the counting function at the positions of the
real roots/holes. 

We remark that there exists a useful relation between the number of different
types of sources, which in the continuum looks like \cite{ddv-97}

\begin{equation}
\label{conteggi}
N_{H}-2N_{S}=2S+M_{C}+2\theta (\pi -2\gamma )M_{W}\: ,
\end{equation}
which we call the \emph{counting equation.} The analogous relation on the lattice
differs by a contribution from sources that disappear in the continuum limit.
One can define 
\[
N_{H,eff}=N_{H}-2N_{S}\: ,\]
which is in fact the total number of the physical solitonic particles (i.e.
solitons and antisolitons) in the given state and is independent of \( l \), which
we proceed to explain. Suppose that \( I \) is a Bethe quantum number corresponding
to some special objects and let us denote by \( \{z_{j}\} \) the set of real numbers for which
\( Z\left( z_{j}\right) =2\pi I \). First note that since \( Z \) is globally increasing the number of these positions
is always odd. For large enough \( l \) the set \( \{z_{j}\} \) has only one element \( z_{1} \), which is
normal but can change its nature when decreasing \( l \) together with the appearance
of new positions \( \{z_{j}:\: j\neq 1\} \). All the new positions which are generated in this way are
holes and satisfy the following properties (for illustration see Figure \ref{spec_exp}):\begin{figure}
{\centering \begin{tabular}{ccc}
\includegraphics{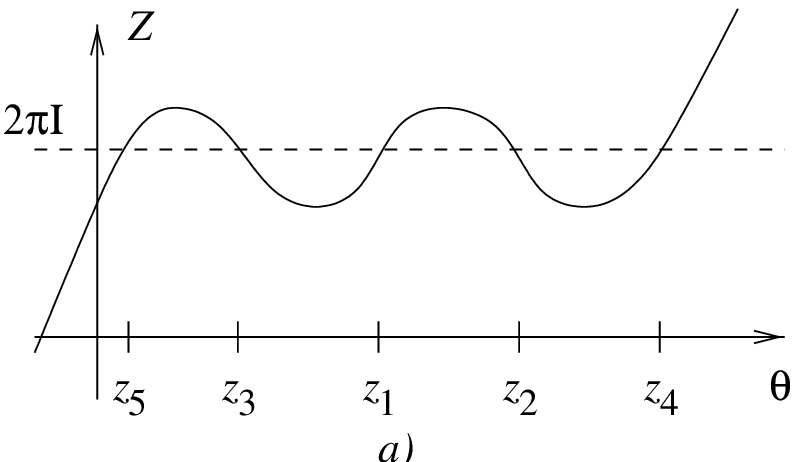} &
&
\includegraphics{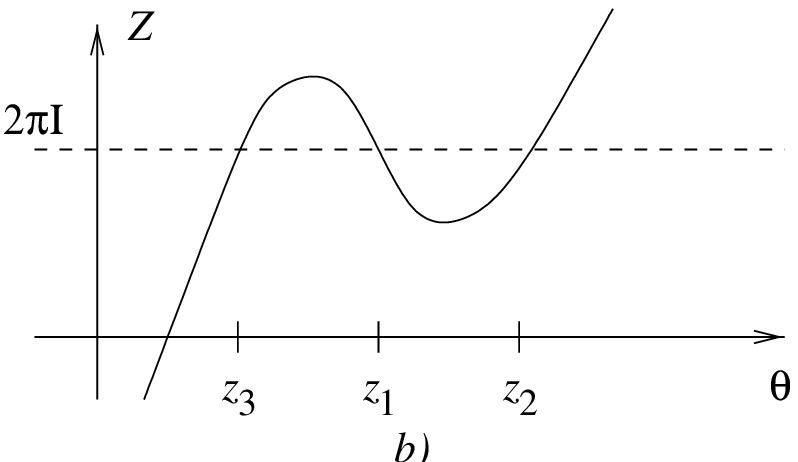} \\
\end{tabular}\par}

\caption{\small \label{spec_exp} The typical behaviour of the counting function \protect\( Z\protect \) when a) \protect\( z_{1}\protect \) is normal and
when b) \protect\( z_{1}\protect \) is special.}
\end{figure} 

\begin{enumerate}
\item If for a given \( l \) \( z_{1} \) is normal then half of the other \( \{z_{j}:\: j\neq 1\} \) are special, while the
other half is normal. For that reason the part they contribute to \( N_{H} \) is the double
of their contribution to \( N_{S} \) and so they do not change \( N_{H,eff} \). Note that special holes
have to be counted both as special objects and as holes (and therefore contribute
to the source term with two pieces).
\item If \( z_{1} \) is special, then the number \( \tilde{N}_{S} \) of the special objects in \( \{z_{j}:\: j\neq 1\} \) is related to
the number \( \tilde{N}_{H} \) of holes in the same set by \( \tilde{N}_{H}=2(\tilde{N}_{S}+1). \) Together with the assumption that
\( z_{1} \) is special we obtain a vanishing contribution to \( N_{H,eff} \) again.
\end{enumerate}

\section{\label{IR_limit}Scaling functions in the infrared limit}

\setcounter{equation}{0}

\noindent In this section we analyze the scaling functions coming from the NLIE
in the IR limit. For \( l\rightarrow \infty  \) the convolution term in the equation becomes small (of
order \( O\left( e^{-l}\right)  \)) and can be dropped. The surviving part can be seen as a dressed higher
level Bethe Ansatz giving constraints on the asymptotic states. In particular,
the scattering physics can be reproduced in this limit and the S-matrix can
be read out of the NLIE. This analysis was performed in \cite{ddv-97} for pure hole states;
here we extend it in order to extract results on the quantisation rules of complex
roots which will be crucial in the sequel.

\subsection{States with holes only}

As our first example, let us take states with only holes in the sea of real
Bethe Ansatz roots. In the IR limit we can write

\begin{equation}
\label{purehole_BA}
Z(\vartheta )=l\sinh \vartheta +\displaystyle\sum _{j=1}^{N_{H}}\chi (\vartheta -h_{j})\quad ,\quad Z(h_{j})=2\pi I_{j}.
\end{equation}
Since the function \( \chi  \) can be written as

\begin{equation}
\label{phaseshift}
\chi (\vartheta )=-i\log S_{++}^{++}(\vartheta )\: ,
\end{equation}
where \( S_{++}^{++}(\vartheta ) \) is the soliton-soliton scattering amplitude in sine-Gordon theory, the
equations (\ref{purehole_BA}) describe the (approximate) dressed Bethe Ansatz equations for
states containing only solitons (or antisolitons). The IR limit puts no restriction
on the quantum numbers \( I_{k} \) of the holes; this seems to be a very general feature
that remains true even when complex roots are allowed.

\subsection{\label{IRlimit_neutral}Neutral two-particle states}

Let us now extend our investigation to situations with complex roots and consider
the two particle states in more detail. Forgetting for the moment the breathers,
we have to consider the two-soliton states. The soliton-antisoliton come in
doublets so there are four different polarizations for two particle states of
which the \( ss \) and \( \bar{s}\bar{s} \) have topological charge \( Q=\pm 2 \). These two states are expected to
have exactly the same scaling function for energy and momentum because the sG/mTh
theory is charge conjugation invariant. There are two different situations for
the neutral \( s\bar{s} \) state, which have spatially symmetric and antisymmetric wavefunctions
(denoted by \( (s\bar{s})_{+} \) and \( (s\bar{s})_{-} \), respectively). To separate the symmetric and antisymmetric
part one simply has to diagonalize the \( 4\times 4 \) SG two particle S-matrix and see that
it has 2 coinciding eigenvalues (equal to \( e^{i\chi (\vartheta )} \), corresponding to \( ss \) and \( \bar{s}\bar{s} \)), and
two different eigenvalues in the \( Q=0 \) channel.

Now we proceed to demonstrate that the IR limit restricts the possible quantum
numbers of the complex roots. To simplify matters we consider only the repulsive
regime \( p>1 \).

In the repulsive regime \( p>1 \), the scaling function \( (s\bar{s})_{-} \) is realized as the solution
to the NLIE with two holes (at positions \( \vartheta _{1,2} \)) and a complex pair at the position
\( \rho \pm i\sigma  \) . In the IR limit we have

\noindent 
\begin{equation}
\label{}
\begin{array}{l}
\begin{array}{lcl}
Z(\vartheta ) & = & l\sinh \vartheta +\chi (\vartheta -\vartheta _{1})+\chi (\vartheta -\vartheta _{2})-\chi (\vartheta -\rho -i\sigma )-\chi (\vartheta -\rho +i\sigma ),\\
 & 
\end{array}\\
Z(\vartheta _{1,2})=2\pi I_{1,2}\; ,\\
Z(\rho \pm i\sigma )=2\pi I^{\pm }_{c}\; .
\end{array}
\end{equation}
The quantization condition for the complex roots explicitly reads (we write
down the equation only for the upper member of the complex pair, since the other
one is similar)

\begin{equation}
\label{complex_quant}
l\sinh (\rho +i\sigma )+\chi (\rho +i\sigma -\vartheta _{1})+\chi (\rho +i\sigma -\vartheta _{2})-\chi (2i\sigma )=2\pi I^{+}_{c\: .}
\end{equation}
Now observe that as \( l\: \rightarrow \: \infty  \), the first term on the left hand side acquires a large
imaginary part, but the right hand side is strictly real. The imaginary contribution
should be cancelled by some other term. The function \( \chi  \) is bounded everywhere
except for isolated logarithmic singularities on the imaginary axis. For the
cancellation of the imaginary part the argument \( 2i\sigma  \) of the last term on the left
hand side should approach one of these singularities (similarly to the analysis
in TBA \cite{tateo_dorey}). In the repulsive regime, taking into account that for a close pair
\( \sigma <\pi  \), the only possible choice for \( \sigma  \) is to approach \( \displaystyle\frac{\pi }{2} \) as \( l\: \rightarrow \: \infty  \). The soliton-soliton
scattering amplitude has a simple zero at \( \vartheta =i\pi  \) with a derivative which we denote
by \( C \) (the exact value does not matter.) To leading order in \( l \), the cancellation
of the imaginary part reads

\begin{equation}
\label{}
l\cosh \rho +\Re e\log C\left( \sigma -\displaystyle\frac{\pi }{2}\right) =0\: ,
\end{equation}
from which we deduce

\begin{equation}
\label{complex_as}
\left| \sigma -\displaystyle\frac{\pi }{2}\right| \sim \exp \left( -l\cosh \left( \rho \right) \right) \: ,
\end{equation}
so the imaginary part of the complex pair approaches its infrared limit exponentially
fast. This approach is modified by taking into account the finite imaginary
contributions coming from the source terms of the holes and from the convolution
term. These contributions lead to corrections of the order \( e^{-l} \) and so they modify
only the value of the constant \( C \). 

For the real part we get, again to the leading order

\begin{equation}
\label{}
\Re e\chi \left( \rho +i\displaystyle\frac{\pi }{2}-\vartheta _{1}\right) +\Re e\chi \left( \rho +i\displaystyle\frac{\pi }{2}-\vartheta _{2}\right) =2\pi I^{+}_{c\: .}
\end{equation}
It can be shown that (in the repulsive regime \( p>1 \))

\begin{equation}
\label{}
\xi \left( \vartheta \right) \equiv \Re e\chi \left( \vartheta +i\displaystyle\frac{\pi }{2}\right) =-\displaystyle\frac{i}{2}\log \displaystyle\frac{\sinh \displaystyle\frac{1}{p}\left( i\displaystyle\frac{\pi }{2}-\vartheta \right) }{\sinh \displaystyle\frac{1}{p}\left( i\displaystyle\frac{\pi }{2}+\vartheta \right) }\: ,
\end{equation}
where the branch of the logarithm is specified by \( \xi \left( 0\right) =0 \) and continuity. \( \xi  \) is an
odd monotonous function bounded by

\begin{equation}
\label{}
\left| \xi (\vartheta )\right| \leq \left| \xi (\infty )\right| =\displaystyle\frac{\pi (p-1)}{2p}
\end{equation}
from below and above. This means that for any allowed value of \( I^{\pm }_{c} \) the real position
of the complex pair is determined uniquely and that

\begin{equation}
\label{}
\left| I^{\pm }_{c}\right| <\left| \displaystyle\frac{p-1}{2p}\right| \: ,
\end{equation}
and since in the repulsive regime \( p>1 \), the only possible choice is \( I^{\pm }_{c}=0 \). Then the
solution for \( \rho  \) is

\begin{equation}
\label{}
\rho =\displaystyle\frac{\vartheta _{1}+\vartheta _{2}}{2}\: ,
\end{equation}
so it approaches the central position between the two holes (the corrections
to this asymptotics are also exponentially small for large \( l \)). In fact, for
the symmetric hole configuration \( I_{1}=-I_{2} \) we expect \( \vartheta _{1}=-\vartheta _{2} \) and \( \rho =0 \) to be valid even for finite
values \( l \). 

\begin{figure}
{\centering \resizebox*{0.6\textwidth}{0.2\textheight}{\includegraphics{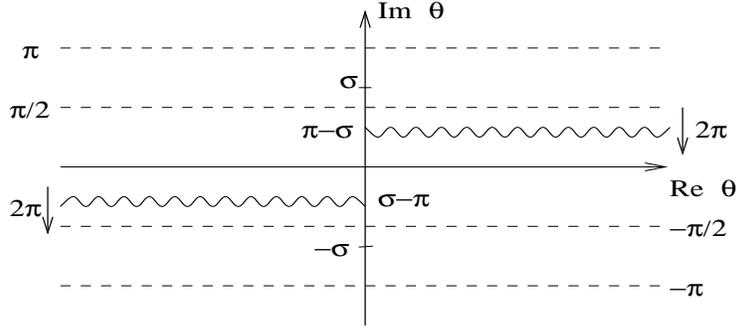}} \par}

\caption{\small \label{cuts_figure} The analytic structure of the counting function if \protect\( \sigma >\frac{\pi }{2}\protect \). The logarithmic cuts
are indicated with the wiggly lines. The one lying in the upper half plane originates
from the root in the lower half plane and vice versa. We also indicated the
value of the discontinuity across the cuts.}
\end{figure}

However, the above derivation is valid only if \( \sigma <\displaystyle\frac{\pi }{2} \) and so we do not cross the
boundary of the analyticity strip of the \( \chi (\vartheta ) \) function which is at \( \Im m\, \, \vartheta =\pi  \). If \( \sigma >\displaystyle\frac{\pi }{2} \) we have
to use the second determination, which leads us to a different conclusion. Equation
(\ref{complex_quant}) can be written in the form
\[
l\sinh (\rho +i\sigma )+\chi (\rho +i\sigma -\vartheta _{1})+\chi (\rho +i\sigma -\vartheta _{2})-\chi (2i\sigma )_{II}+\chi (2i\sigma -i\pi )=2\pi I_{C}\, \, ,\]
where
\[
\chi (\vartheta )_{II}=\chi (\vartheta )+\chi (\vartheta -i\pi )=-i\log \frac{\sinh \left( \frac{i\pi -\vartheta }{p}\right) }{\sinh \left( \frac{\vartheta }{p}\right) }\, \, .\]
The conclusion that \( \sigma  \) approaches \( \frac{\pi }{2} \) exponentially fast as \( l\, \rightarrow \, \infty  \) is unchanged, but
now the real part of \( \chi (2i\sigma ) \) is not zero but instead \( \pm \pi  \) (modulo \( 2\pi  \)). We choose the branches
of the logarithm in such a way that \( \Re e\, \chi (\pm 2i\sigma )=\pm \pi  \) (the structure of the cuts is displayed
in figure \ref{cuts_figure}). In this way we obtain that 
\[
I^{\pm }_{c}=\mp \frac{1}{2}\, \, ,\]
and the asymptotic value of \( \rho  \) is just as before.

The S-matrix of the \( (s\overline{s})_{-} \) configuration can be obtained by substituting the asymptotic
values

\begin{equation}
\label{}
\rho =\displaystyle\frac{\vartheta _{1}+\vartheta _{2}}{2}\: ,\: \sigma =i\displaystyle\frac{\pi }{2}
\end{equation}
into the expression for \( Z(\vartheta ) \) and considering now the quantisation rules of the
holes.We obtain

\begin{equation}
\label{IRBA_minus}
Z(\vartheta _{1})=l\sinh \left( \vartheta _{1}\right) -i\log S_{-}(\vartheta _{1}-\vartheta _{2})=2\pi I_{1}\: ,
\end{equation}
and a similar equation for the second hole, with

\begin{equation}
\label{}
S_{-}(\vartheta )=-\displaystyle\frac{\sinh \left( \displaystyle\frac{\vartheta +i\pi }{2p}\right) }{\sinh \left( \displaystyle\frac{\vartheta -i\pi }{2p}\right) }S^{++}_{++}(\vartheta )\: ,
\end{equation}
which is the correct scattering amplitude for the antisymmetric configuration
of the soliton-antisoliton system. The amplitude has poles at \( \vartheta =i\pi (1-2kp)\, ,\, k=1,2,\ldots  \) , corresponding
to breathers with mass

\[
m_{2k}=2{\cal M}\sin kp\pi \: .\]

The equation (\ref{IRBA_minus}) describes an approximation to the full NLIE valid for large
\( l \) which is called the dressed or asymptotic Bethe Ansatz. There are three levels
of approximations to the scaling functions: the full NLIE (which is in fact
exact), the higher level Bethe Ansatz (HLBA) obtained from the NLIE by dropping
the convolution term and the asymptotic BA. The difference between the BA and
the HLBA is that while the former keeps the complex pair in its asymptotic position,
the latter takes into account the corrections coming from the fact that \( \sigma  \) changes
with \( l \) (to first order as given in (\ref{complex_as}) ). However, since the convolution term
is of the same order in \( l \) as the dependence of \( \sigma  \) derived from the HLBA, the
HLBA is not a self-consistent approximation scheme. Therefore we are left with
the exact NLIE, which is valid for all scales and with the BA as its IR asymptotic
form.

The above conclusions about the \( (s\bar{s})_{-} \) state an be extended into the attractive regime
as long as \( p>\frac{1}{2} \). At the point \( p=\frac{1}{2} \), however, the pair situated at the asymptotic position
\( \pm i\frac{\pi }{2} \) hits the boundary of the fundamental analyticity strip situated at \( i\pi p \). This
phenomenon has a physical origin: this point is the threshold for the second
breather bound state which is the first pole entering the physical strip in
the \( (s\bar{s})_{-} \) channel. To continue our state further, it requires to go to configurations
having an array of complex roots of the first type (cf. \cite{ddv-97}). Without going into
further details, let us mention only that any such array contains a close pair
plus some wide pairs depending on the value of \( p \). We return to comment on this
later in subsection \ref{example_2handcp}. 

The symmetric state \( (s\bar{s})_{+} \) can be obtained from a configuration with two holes and
one selfconjugate wide root. Considerations similar to the above lead to the
scattering amplitude

\begin{equation}
\label{}
S_{+}(\vartheta )=\displaystyle\frac{\cosh \left( \displaystyle\frac{\vartheta +i\pi }{2p}\right) }{\cosh \left( \displaystyle\frac{\vartheta -i\pi }{2p}\right) }S^{++}_{++}(\theta )\: ,
\end{equation}
which agrees with the prediction from the exact S-matrix. The poles are now
at \( \vartheta =i\pi (1-(2k+1)p)\, ,\, k=0,1,\ldots  \) and correspond to breathers with mass
\[
m_{2k+1}=2{\cal M}\sin \frac{(2k+1)p\pi }{2}\: .\]
This configuration extends down to \( p=1 \), where the selfconjugate root collides
with the boundary of the fundamental analyticity strip at \( i\pi  \). This is exactly
the threshold for the first breather which is the lowest bound state in the
\( (s\bar{s})_{+} \) channel. For \( p<1 \) the \( (s\bar{s})_{+} \) state contains a degenerate array of the first kind (cf.
\cite{ddv-97}), which consists of a close pair, a selfconjugate root and some wide pairs,
again depending on the value of \( p \).

We remark that it is easy to extend the above considerations to a state with
four holes and a complex pair. In this case the only essential modification
to the above conclusions is that the limit for the complex pair quantum number
becomes
\begin{equation}
\label{}
\left| I^{\pm }_{c}\right| <\left| \displaystyle\frac{p-1}{p}\right| 
\end{equation}
when \( \sigma <\displaystyle\frac{\pi }{2} \), and
\begin{equation}
\label{}
\left| I^{\pm }_{c}\pm \displaystyle\frac{1}{2}\right| <\left| \displaystyle\frac{p-1}{p}\right| \: ,
\end{equation}
when \( \sigma >\displaystyle\frac{\pi }{2} \), since now there are four \( \chi  \) sources coming from the holes. In the range
\( 1<p<2 \) this gives the same constraints
\[
I^{\pm }_{c}=\left\{ \begin{array}{ll}
0\, , & \, \, \sigma <\displaystyle\frac{\pi }{2}\, \, ,\\
\mp \displaystyle\frac{1}{2}\, , & \, \, \sigma >\displaystyle\frac{\pi }{2}
\end{array}\right. \, \, ,\]
as for the \( (s\bar{s})_{-} \) state (for \( p>2 \) it allows for more solutions, which however we will
not need in the sequel).

\section{\label{UV_limit}UV limit and kink approximation}

\setcounter{equation}{0}

\subsection{Calculation of the ultraviolet conformal weights}

We now examine the ultraviolet limit of the states described by the NLIE in
order to compare it with the known facts about the UV limit of the sG/mTh theory,
outlined in subsection \ref{sG_mTh_relation}. We follow closely the approach described in detail
in \cite{ddv-97}. However, our NLIE differs from the one proposed in \cite{ddv-97} as we explained in
section \ref{NLIE_derivation}. Furthermore, we do not fix \( \delta  \) to be \( S\bmod 2 \) as the authors of \cite{ddv-97} do but instead
leave it as a free parameter taking the values \( 0 \) and \( 1 \). For these reasons, it
is worthwhile to review the main points of the derivation. For simplicity we
first suppose that there are no selfconjugate wide roots, for which we had to
redefine the source term in the repulsive regime (see section \ref{NLIE_derivation}).

From conformal perturbation it is known that the leading ultraviolet behaviour
of energy and momentum are given by (c.f. (\ref{scalingfun})):
\begin{equation}
\label{andamento-uv}
\begin{array}{rl}
\displaystyle E(L)= & -\displaystyle\frac{\pi }{6L}\left( c-12(\Delta _{+}+\Delta _{-})\right) +\ldots \: ,\\
P(L)= & \displaystyle\frac{2\pi }{L}\left( \Delta _{+}-\Delta _{-}\right) +\ldots \: .
\end{array}
\end{equation}
Therefore, for the UV comparison we have to determine the contributions to the
energy and momentum which behave as \( 1/L \) when \( L\rightarrow 0 \). Consider e.g. the contribution
of a hole positioned at \( h \) to the energy which is \( m\cosh h \). To get a \( 1/L \) contribution
in the UV limit we have to require
\begin{equation}
\label{log-uv}
h=finite\pm \log \displaystyle\frac{2}{l}\: ,
\end{equation}
which means that we are principally interested in the holes whose position has
a logarithmically diverging term as \( l\rightarrow 0 \). Similar argument can be applied to complex
roots.

The behaviour of the sources for \( l\rightarrow 0 \) can be classified into three possibilities:
their position can remain finite (we call them ``central''), or they can move
towards the two infinities as \( \pm \log \displaystyle\frac{2}{l} \) (``left/right movers''). For sources in all three
classes we can introduce the finite parts \( \vartheta ^{\pm ,0}_{j} \) of their positions \( \vartheta _{j} \) by extracting
the divergent contribution:
\begin{equation}
\label{finite_parts}
\left\{ \vartheta _{j}\right\} \, \rightarrow \, \left\{ \vartheta _{j}^{\pm }\pm \log \displaystyle\frac{2}{l},\, \vartheta _{j}^{0}\right\} \: .
\end{equation}
We denote the number of right/left moving resp. central holes by \( N^{\pm ,0}_{H} \) and similarly
we introduce the numbers \( N^{\pm ,0}_{S} \), \( M^{\pm ,0}_{C} \) and \( M^{\pm ,0}_{W} \).

In the UV limit it is expected that the NLIE can be split into three separate
equations for the three asymptotic regions, which are separated by \( \log \displaystyle\frac{2}{l} \). To derive
these equations, observe that \( Z \) has an implicit dependence from \( l \) which we make
manifest writing \( Z=Z(\vartheta ,\: l). \) Let us define the so-called ``kink functions'' in the following
way:
\begin{equation}
\label{kink_functions}
Z_{\pm }(\vartheta )=\lim _{l\rightarrow 0}Z\left( \vartheta \pm \log \displaystyle\frac{2}{l},\: l\right) \, .
\end{equation}
The term ``kink'' has its origin in the fact that the asymptotic form of the function
\( Z(\vartheta ) \) has two plateaus stretching between the central region and the regions of
the left/right movers. The functions \( Z_{\pm }(\vartheta ) \) describe the interpolation between the
plateaus and the asymptotic behaviour of \( Z(\vartheta ) \) at the corresponding infinity. Note
that if there are no central objects then the two plateaus merge into a single
one stretching from the left movers' region to the right movers' one.

We introduce the quantities \( S^{+},\: S^{-},\: S^{0} \) by the definition: 
\[
S^{\pm ,0}=\frac{1}{2}[N_{H}^{\pm ,0}-2N_{S}^{\pm ,0}-M_{C}^{\pm ,0}-2M^{\pm ,0}_{W}\theta (\pi -2\gamma )]\]
in analogy with the counting equation (\ref{conteggi}) for \( S, \) except for the fact that \( S \) is
always integer, whereas \( S^{\pm } \) can be half-integer; moreover, in some cases, \( S^{0} \) can
be negative. They are interpreted as the spin of the right/left moving and central
solutions. It is obvious that
\[
S=S^{+}+S^{-}+S^{0}.\]
We also denote 
\[
\displaystyle \chi _{\infty }=\chi (+\infty )=\pi \frac{\pi /2-\gamma }{\pi -\gamma }\: .\]
For convenience we switch to \( \gamma  \) as the coupling parameter in our UV calculations.
Substituting the expressions (\ref{finite_parts}) and (\ref{kink_functions}) into the NLIE (\ref{nlie-cont}) and keeping only
the leading terms as \( l\rightarrow 0 \) we obtain for the left/right movers:
\begin{equation}
\label{kink}
\displaystyle Z_{\pm }(\vartheta )=\pm e^{\pm \vartheta }+g_{\pm }(\vartheta )+\displaystyle\int\limits ^{\infty }_{-\infty }dxG(\vartheta -x){\cal Q}_{\pm }(x)\: ,
\end{equation}
where the source term \( g_{\pm }(\vartheta ) \) is given by
\begin{equation}
\label{kink_source}
\begin{array}{rl}
\displaystyle g_{\pm }(\vartheta )= & \pm 2\chi _{\infty }\left( S-S^{\pm }\right) +2\pi l_{W}^{\pm }+\displaystyle\sum ^{N^{\pm }_{H}}_{j=1}\chi \left( \vartheta -h_{j}^{\pm }\right) -\\
 & 2\displaystyle\sum ^{N^{\pm }_{S}}_{j=1}\chi \left( \vartheta -y_{j}^{\pm }\right) -\displaystyle\sum ^{M_{C}^{\pm }}_{j=1}\chi \left( \vartheta -c_{j}^{\pm }\right) -\displaystyle\sum ^{M^{\pm }_{W}}_{j=1}\chi \left( \vartheta -w^{\pm }_{j}\right) _{II}\: ,
\end{array}
\end{equation}
where \( l^{\pm }_{W} \) is an integer depending on the configuration of the wide roots (and
also on the regime of coupling taken), whose explicit expression is not important
in the sequel. The quantization conditions read
\[
Z_{\pm }\left( \vartheta _{j}^{\pm }\right) =2\pi I_{j}^{\pm }\, .\]
The equation for the central region can be obtained by taking \( l\rightarrow 0 \) in (\ref{nlie-cont}):
\[
\displaystyle Z_{0}(\vartheta )=g_{0}(\vartheta )+\int\limits ^{\infty }_{-\infty }dxG(\vartheta -x){\cal Q}_{0}(x)\: ,\]
together with
\[
\begin{array}{rl}
\displaystyle g_{0}(\vartheta )= & 2\chi _{\infty }\left( S^{-}-S^{+}\right) +2\pi l^{0}_{W}+\displaystyle\sum ^{N^{0}_{H}}_{j=1}\chi \left( \vartheta -h^{0}_{j}\right) -\\
 & 2\displaystyle\sum ^{N^{0}_{S}}_{j=1}\chi \left( \vartheta -y^{0}_{j}\right) -\displaystyle\sum ^{M_{C}^{0}}_{j=1}\chi \left( \vartheta -c^{0}_{j}\right) -\displaystyle\sum ^{M^{0}_{W}}_{j=1}\chi _{II}\left( \vartheta -w^{0}_{j}\right) \, ,
\end{array}\]

\[
Z_{0}\left( \vartheta _{j}^{0}\right) =2\pi I_{j}^{0}\, ,\]
where \( Z_{0}(\lambda )=\lim _{l\rightarrow 0}Z(\lambda ,l) \). 

The functions \( {\cal Q}_{\pm ,0}(x) \) are connected with the corresponding \( Z_{\pm ,0}(x) \) by the usual expression
(\ref{Q_Z_relation}):
\[
{\cal Q}_{\sigma }(x)=\frac{1}{i}\log \frac{1+(-1)^{\delta }e^{iZ_{\sigma }(x+i\eta )}}{1+(-1)^{\delta }e^{-iZ_{\sigma }(x-i\eta )}}\, ,\quad \; -\pi <{\cal Q}_{\sigma }(x)\leq \pi \, ,\]
where \( \sigma =\pm ,0 \) and \( \eta  \) takes an infinitesimal positive real value.

In the following we will often need the asymptotic values of \( Z_{\sigma } \) and \( {\cal Q}_{\sigma } \)\( . \) From (\ref{kink})
we obtain for \( Z_{\pm } \): 
\begin{equation}
\label{zeta-infty}
Z_{+}(+\infty )=+\infty =-Z_{-}(-\infty )\, ,
\end{equation}
the corresponding asymptotic value for \( {\cal Q}_{\pm } \) is: 
\begin{equation}
\label{q-infty}
{\cal Q}_{\pm }(\pm \infty )=0\, .
\end{equation}
For the other limiting values of \( Z_{\pm } \) we obtain from (\ref{kink}) the plateau equation:
\begin{equation}
\label{plateau}
Z_{\pm }(\mp \infty )=g_{\pm }(\mp \infty )+\displaystyle\frac{\chi _{\infty }}{\pi }{\cal Q}_{\pm }(\mp \infty )\, ,
\end{equation}
while from the definition of \( {\cal Q} \) we have
\[
Z_{\sigma }(\vartheta )={\cal Q}_{\sigma }(\vartheta )+\pi \delta +2\pi k\: .\]
From (\ref{kink_source}) we obtain
\[
g_{\pm }(\mp \infty )=\pm 2\left( S-2S^{\pm }\right) \chi _{\infty }+2\pi k^{\pm }_{W}\: ,\]
where again \( k^{\pm }_{W} \) are some integers depending on wide roots. This leads to 
\begin{equation}
\label{plateau_solution}
{\cal Q}_{\pm }(\mp \infty )=\pm 2(\pi -2\gamma )\left( S-2S^{\pm }\right) -2(\pi -\gamma )(\delta +2k_{\pm })\, ,
\end{equation}
\( k_{\pm } \) are integers determined by the condition \( -\pi <{\cal Q}_{+}(-\infty )\leq \pi  \). Its presence allows us to absorb
the integers \( k^{\pm }_{W} \). Including selfconjugate wide roots means that \( k^{\pm }_{W} \) can take half-integer
values. By a long but elementary analysis one can see that either both of them
are integer or half-integer depending on whether the total number of selfconjugate
roots is even or odd, respectively. This means that in the context of the UV
analysis we can define an effective value of \( \delta  \) 
\[
\delta _{eff}=\delta +(2k^{\pm }_{W}\bmod 2)=\delta +(number\: of\: self-conjugate\: roots\: \bmod 2)\: .\]
In the repulsive regime there is no ambiguity in the choice of \( k \) because \( 4(\pi -\gamma )\geq 2\pi  \),
which is larger than the width of the fundamental strip of the logarithm\( . \) In
some cases compatible with the counting equation apparently there is no solution
for the plateau equation (\ref{plateau}). This problem can be overcome by observing that
some special sources (roots/holes) are required as we will demonstrate on explicit
examples later.

Considering now the region \( \gamma >\pi /2 \), there can be more than one different possibilities
to choose \( k \). In \cite{ddv-97} it is shown that this phenomenon is related to the fact that
wide roots become excitations independent of the holes, which is manifestly
evident from the fact that they no longer contribute to the spin of the state
\( S \) and that the asymptotic value of their source contribution \( \chi _{II} \) in the attractive
regime is \( 0 \) modulo \( 2\pi  \).

For the central part we obtain the following equations:
\[
Z_{0}(\pm \infty )=\pm 2(S-2S^{\pm })\chi _{\infty }+2\pi k^{\pm }_{W}+\frac{\chi _{\infty }}{\pi }{\cal Q}_{0}(\pm \infty )\, ,\]
that coincide respectively with equations for \( {\cal Q}_{+}(-\infty ) \) and \( {\cal Q}_{-}(+\infty ). \) This implies that 
\[
{\cal Q}_{0}(+\infty )={\cal Q}_{+}(-\infty )\quad ,\quad {\cal Q}_{0}(-\infty )={\cal Q}_{-}(+\infty )\, ,\]
corresponding to the formation of the plateau regions as described above.

Now we are ready to perform the ultraviolet limit on the expressions of the
energy and momentum. To this end we substitute the ultraviolet behaviour of
the sources into the energy and momentum expressions and retain only terms of
order \( \frac{1}{L}. \) The kinetic terms give contributions that look like
\begin{equation}
\label{source}
\displaystyle\frac{1}{L}e^{\pm \vartheta _{k}^{\pm }}.
\end{equation}
 The integral terms must be calculated separately. Their form in the energy
and momentum expressions is, respectively:
\[
\begin{array}{c}
\displaystyle -{\cal M}\displaystyle\int\limits ^{\infty }_{-\infty }\displaystyle\frac{dx}{2\pi }\sinh x\, {\cal Q}(x)=-{\cal M}\displaystyle\int\limits ^{\infty }_{-\infty }\displaystyle\frac{dx}{2\pi }\displaystyle\frac{1}{2}[e^{x}\, {\cal Q}(x)-e^{-x}\, {\cal Q}(x)]\, ,\\
-{\cal M}\displaystyle\int\limits _{-\infty }^{\infty }\displaystyle\frac{dx}{2\pi }\cosh x\, {\cal Q}(x)=-{\cal M}\displaystyle\int\limits _{-\infty }^{\infty }\displaystyle\frac{dx}{2\pi }\displaystyle\frac{1}{2}[e^{x}\, {\cal Q}(x)+e^{-x}\, {\cal Q}(x)]\, ,
\end{array}\]
where the contribution depending on \( e^{x} \) is called \emph{``+'' kink contribution},
and the other term is the \emph{``-'' kink contribution}. For the computation
observe that, in the limit of very small \( l \), it is possible to write an expression
for \( {\cal Q} \):
\begin{equation}
\label{qu-asintotico}
{\cal Q}(x,l)\sim {\cal Q}_{-}\left( x+\log \displaystyle\frac{2}{l}\right) +{\cal Q}_{+}\left( x-\log \displaystyle\frac{2}{l}\right) -{\cal Q}_{0}(-x)+q(x,l)\: ,
\end{equation}
where the function \( q(x,l) \) vanishes in the \( l\rightarrow 0 \) limit.  

Substituting (\ref{qu-asintotico}) in the integral form of the ``+'' kink contribution we obtain

\begin{equation}
\label{kink+}
{\cal M}\displaystyle\int\limits ^{\infty }_{-\infty }\displaystyle\frac{dx}{2\pi }\displaystyle\frac{1}{2}e^{x}\, {\cal Q}(x)\doteq \displaystyle\frac{1}{L}\displaystyle\int\limits ^{\infty }_{-\infty }\displaystyle\frac{dx}{2\pi }e^{x}\, {\cal Q}_{+}(x)\, ,
\end{equation}
where the symbol \( \doteq  \) means that only the terms in \( 1/L \) are retained. Similarly we
get for the ``-'' kink term
\begin{equation}
\label{kink-}
{\cal M}\displaystyle\int\limits ^{\infty }_{-\infty }\displaystyle\frac{dx}{2\pi }\displaystyle\frac{1}{2}e^{-x}\, {\cal Q}(x)\doteq \displaystyle\frac{1}{L}\displaystyle\int\limits ^{\infty }_{-\infty }\displaystyle\frac{dx}{2\pi }e^{-x}\, {\cal Q}_{-}(x)\, .
\end{equation}
Now we are ready to express energy and momentum in a way dependent only on quantities
which are finite in the UV limit. For the conformal energy (scaling weight)
we obtain

\[
\begin{array}{rl}
\displaystyle \Delta +\bar{\Delta }= & \displaystyle\frac{c}{12}+\displaystyle\frac{1}{2\pi }\displaystyle\sum _{\sigma =\pm }\Bigg (\displaystyle\sum ^{N^{\sigma }_{H}}_{j=1}e^{\sigma h^{\sigma }_{j}}-2\displaystyle\sum ^{N^{\sigma }_{S}}_{j=1}e^{\sigma y^{\sigma }_{j}}-\displaystyle\sum ^{M^{\sigma }_{C}}_{j=1}e^{\sigma c^{\sigma }_{j}}+\displaystyle\sum ^{M^{\sigma }_{W}}_{j=1}\left( e^{\sigma w_{j}^{\sigma }}\right) _{II}\\
 & -\displaystyle\int\limits ^{\infty }_{-\infty }\displaystyle\frac{dx}{2\pi }\sigma e^{\sigma x}{\cal Q}_{\sigma }(x)\Bigg )\: ,
\end{array}\]
and the conformal spin is given by

\[
\begin{array}{rl}
\displaystyle \Delta -\bar{\Delta }= & \displaystyle\frac{1}{2\pi }\displaystyle\sum _{\sigma =\pm }\Bigg (\displaystyle\sum ^{N^{\sigma }_{H}}_{j=1}\sigma e^{\sigma h^{\sigma }_{j}}-2\displaystyle\sum ^{N^{\sigma }_{S}}_{j=1}\sigma e^{\sigma y^{\sigma }_{j}}-\displaystyle\sum ^{M^{\sigma }_{C}}_{j=1}\sigma e^{\sigma c^{\sigma }_{j}}+\displaystyle\sum ^{M^{\sigma }_{W}}_{j=1}\sigma \left( e^{\sigma w_{j}^{\sigma }}\right) _{II}\\
 & -\displaystyle\int\limits ^{\infty }_{-\infty }\displaystyle\frac{dx}{2\pi }e^{\sigma x}{\cal Q}_{\sigma }(x)\Bigg )\: .
\end{array}\]
Using the method explained in \cite{ddv-97}, it is possible to give a closed expression
for the sums and the integrals. The computation is simple, and we perform it
explicitly only for the case of holes. The quantisation condition for a hole
\( h_{j}^{\pm } \) , using the kink equation (\ref{kink}) and the oddness of the \( \chi  \) function allows us
to write:
\begin{equation}
\label{limit}
\begin{array}{rl}
\displaystyle \pm \displaystyle\sum ^{N^{\pm }_{H}}_{j=1}e^{\pm h_{j}^{\pm }}= & \displaystyle\sum ^{N^{\pm }_{H}}_{j=1}\Bigg (2\pi I_{h_{j}}^{\pm }+2\displaystyle\sum ^{N^{\pm }_{S}}_{k=1}\chi \left( h^{\pm }_{j}-y^{\pm }_{k}\right) +\displaystyle\sum ^{M^{\pm }_{C}}_{k=1}\chi \left( h^{\pm }_{j}-c^{\pm }_{k}\right) \\
+ & \displaystyle\sum ^{M^{\pm }_{W}}_{k=1}\chi _{II}\left( h^{\pm }_{j}-w^{\pm }_{k}\right) \Bigg )\mp N^{\pm }_{H}\left( 2(S-S^{\pm })\chi _{\infty }+2\pi l^{\pm }_{W}\right) \\
- & \displaystyle\sum ^{N^{\pm }_{H}}_{j=1}\displaystyle\int ^{\infty }_{-\infty }dxG\left( h^{\pm }_{j}-x\right) {\cal Q}_{\pm }(x)\, .
\end{array}
\end{equation}
The contribution from the other sources can be evaluated in a similar way. Summing
together all the expressions obtained in this way the terms containing \( \chi  \) cancel
completely because of the oddness of the function \( \chi  \). In the integral term we
make the substitution 
\begin{equation}
\label{source-uv}
2\pi G(x)=\chi '(x)\, .
\end{equation}
The result is: 
\[
\begin{array}{rl}
\displaystyle \Delta _{\pm } & =\displaystyle\frac{c}{24}\pm \left( I^{\pm }_{H}-2I^{\pm }_{S}-I^{\pm }_{C}-I^{\pm }_{W}\right) +\displaystyle\frac{1}{2\pi }\Sigma _{\pm }\\
 & \mp \displaystyle\frac{1}{2\pi }\displaystyle\int\limits ^{\infty }_{-\infty }\displaystyle\frac{dx}{2\pi }\varphi ^{,}_{\pm }(x){\cal Q}_{\pm }(x)\: ,
\end{array}\]
where
\begin{equation}
\label{var-fi}
\displaystyle \varphi _{\pm }(\vartheta )=Z_{\pm }(\vartheta )-\displaystyle\int ^{\infty }_{-\infty }dxG(\vartheta -x){\cal Q}_{\pm }(x)\: ,
\end{equation}

\begin{equation}
\label{sigmapm}
\Sigma _{\pm }=-4S^{\pm }(S-S^{\pm })\chi _{\infty }+2\pi q^{\pm }_{W}\, ,
\end{equation}
and we denote 
\[
I^{\pm }_{H}=\sum ^{N^{\pm }_{H}}_{j=1}I_{h_{j}}^{\pm }\, ,\, I^{\pm }_{C}=\sum ^{M^{\pm }_{C}}_{j=1}I_{c_{j}}^{\pm }\, ,\, I^{\pm }_{W}=\sum ^{M^{\pm }_{W}}_{j=1}I_{w_{j}}^{\pm }\, \mathrm{and}\, I^{\pm }_{S}=\sum ^{N^{\pm }_{S}}_{j=1}I_{y_{j}}^{\pm }\: .\]
\( q_{W}^{\pm } \) is an integer or half-integer depending on the configuration of the wide roots,
which is best calculated case by case as it has already been noted in \cite{ddv-97}.

Using the ``kink lemma'' proved in \cite{ddv-97}, the integral term can be evaluated:
\[
\displaystyle \int ^{\infty }_{-\infty }\frac{dx}{2\pi }\varphi ^{,}_{\pm }(x){\cal Q}_{\pm }(x)=\pm \left( \frac{\pi }{12}-\frac{{\cal Q}^{2}_{\pm }(\mp \infty )}{8\pi (1-\gamma /\pi )}\right) \: .\]
Our final expression for the conformal weights is
\begin{equation}
\label{delta}
\Delta _{\pm }=\displaystyle\frac{c-1}{24}\pm \left( I^{\pm }_{H}-2I^{\pm }_{S}-I^{\pm }_{C}-I^{\pm }_{W}\right) +\displaystyle\frac{\Sigma _{\pm }}{2\pi }+\displaystyle\frac{{\cal Q}_{\pm }(\mp \infty )^{2}}{16\pi ^{2}(1-\gamma /\pi )}\: .
\end{equation}
Let us discuss certain properties of the formula (\ref{delta}). Using the relation 
\[
1-\frac{\gamma }{\pi }=\frac{1}{2R^{2}}\]
one can prove with simple algebra that the conformal weights have the form
\begin{equation}
\label{conformal}
\Delta _{\pm }=\displaystyle\frac{1}{2}\displaystyle\frac{n^{2}_{\pm }}{R^{2}}+\displaystyle\frac{1}{8}m^{2}R^{2}\pm \displaystyle\frac{n_{\pm }m}{2}+N_{\pm }\, \, ,
\end{equation}
where 
\begin{equation}
\label{final}
\begin{array}{c}
m=2S\, \, ,\\
n_{\pm }=(\displaystyle\frac{\delta }{2}+k_{\pm })\mp (S-2S^{\pm })\, \, ,
\end{array}
\end{equation}
and 
\[
N_{\pm }=\pm (I^{\pm }_{H}-2I^{\pm }_{S}-I^{\pm }_{C}-I^{\pm }_{W})+q^{\pm }_{W}-2(S^{\pm })^{2}\mp 2S^{\pm }(\frac{\delta }{2}+k_{\pm })\, \, .\]
For a consistent interpretation of the weights \( \Delta _{\pm } \) (\ref{conformal}) should have the form (\ref{UV_weights}).
Matching the \( R^{-2} \) terms we get the condition 
\begin{equation}
\label{condition1}
n^{2}_{+}=n_{-}^{2}\, \, .
\end{equation}
The second condition is that one must be able to choose the sign of \( n \) in such
a way that the left and right descendent numbers (obtained by combining \( N_{\pm } \) with
a term depending on \( n \) and \( m \)) are integers. Since the \( R^{-2} \) contribution comes exclusively
from the last term in (\ref{delta}), the one-plateau systems satisfy trivially the condition
(\ref{condition1}), but even in that case it is not clear whether the descendent numbers are
integers. At the time of this writing we cannot exclude the possibility that
there exist some states for which the conformal weights cannot be interpreted
within the framework of \( c=1 \) CFT. In the numerous examples we have calculated so
far we have not found any such source configuration. If it happens for some
configuration of the sources then the corresponding scaling function must be
excluded from the spectrum.

From the second line of (\ref{final}) we obtain a useful relation
\begin{equation}
\label{nmod2}
2n_{\pm }=(\delta +M_{sc})\, \bmod \, 2\, \, \, \, \mathrm{i}.e.\, \, \, \, 2n_{\pm }=\delta _{eff}\, \bmod \, 2\, \, ,
\end{equation}
where we used that \( 2S \) and \( 4S^{\pm } \) are even.

We also drive attention to the fact that we have no proof that the scaling functions
obtained by the method of NLIE span the complete space of states with even topological
charge. This is an extremely difficult problem due to the following two circumstances:
(1) the dependence of the UV conformal weights from the parameters of the source
configuration is rather complicated and (2) to obtain the allowed values of
the complex roots one has to examine the IR limit as in section \ref{IR_limit}. It is also
clear that the same state can be realized by different root configurations depending
on the value of the coupling constant (see e.g. the \( (s\bar{s})_{\pm } \) states in subsection \ref{IRlimit_neutral}).
The investigation of this question is out of the scope of this paper.

Using the relation (\ref{nmod2}) it can be determined what quantisation rule one has to
impose on the real roots/holes if one aims to get solutions that in the UV limit
reproduce the states of the sG/mTh theory. The rule is 
\[
\delta _{eff}=0\, ,\]
which is different from the rule \( \delta \equiv S\bmod 2 \) advocated by Destri and de Vega in \cite{ddv-97}. Their
result is based on the reasoning that in the continuum limit the length of the
spin chain \( N \) goes to infinity, there is no loss generality in taking \( N \) even.
In fact, this assumption fixes the rule \( \delta \equiv S\bmod 2 \). In our approach, however, we do not
fix \( N \) to be even on the lattice, which allows us to leave the parameter \( \delta  \) free
and to fix it later when we examine the UV limit of the states described by
the NLIE. For states without selfconjugate roots our rule simply means that
the real roots/holes are \emph{always quantised by half-integers}. 

An interesting phenomenon is that the conformal weights obtained depend only
on very generic features of the source configuration such as the asymptotics
of the left and right moving sources and the total spin. This means if we have
a certain source configuration, we can add new sources separately to the right
and the left moving part in such a way that they are separately neutral (i.e.
do not change \( S^{+} \) and \( S^{-} \)). In this way we do not change the primary weight, but
generally we increase the term \( I^{\pm }_{H}-2I^{\pm }_{S}-I^{\pm }_{C}-I^{\pm }_{W} \) and so create descendents of the state we started
with. An example: states which have \( S=0 \) and \( S^{\pm }=0 \) are all descendants of the vacuum,
however complicated their actual source configurations are.

In subsection \ref{continuum_limit} it was claimed that the best way to specify a state in the NLIE
framework is to give the corresponding source configuration for large \( l \) since
then the special sources are absent. In the UV asymptotic calculations, however,
we work just at the opposite end of the range of scales. Therefore a different
method to obtain the special sources is necessary. The approach we apply is
to find a regime in the coupling \( \gamma  \) where the plateau equation (\ref{plateau}) admits a solution
without special sources and then determine the solution in other regimes by
analytically continuing the values of \( Z_{\pm }(\mp \infty ) \) in \( \gamma  \). This will allow us to determine
the special sources and will provide a way to fix the ambiguity of the plateau
solution (\ref{plateau_solution}) in the attractive regime as demonstrated on explicit examples in
the following.

Finally, let us remark that putting \( \delta \equiv S\bmod 2 \) our results for the conformal weights
are identical to the ones obtained by Destri and de Vega (formulas (8.21), (8.22)
and (8.23) in \cite{ddv-97}). We now go on to give a detailed examination of states with
simple source configurations in order to clarify the consequences of the general
result (\ref{delta}). Some of the examples correspond to configurations whose scaling
functions will be compared to the TCS data in section \ref{NLIE_and_TCS}.

\subsection{Examples of states without complex roots}

\subsubsection{The vacuum state}

Starting from a lattice with \( 2N \) sites one finds that the antiferromagnetic ground
state which has spin \( S=0 \), when written in terms of Bethe vectors depends on \( M=N \)
roots of the Bethe equations, all of which are real. We expect that this ground
state corresponds to the vacuum of the field theory. However, there are two
possibilities: one for \( N \) even and the other for \( N \) odd, corresponding to \( \delta =0 \) or
\( \delta =1 \). As we show in the sequel, only one of these states can be identified with
the vacuum for a local field theory having a \( c=1 \) UV limiting CFT. The UV dimensions
of the vacuum are \( \Delta ^{\pm }=0 \). Choosing \( \delta =0 \) we obtain 
\[
c=1,\]
as in \cite{ddv-95}. 

For the choice \( \delta =1 \) case we obtain an interesting result. From (\ref{plateau_solution}) we derive 
\begin{equation}
\label{vuoto-}
{\cal Q}_{+}(-\infty )=-2(\pi -\gamma )(1+2k)\, ,
\end{equation}
which admits solution only in the attractive regime when
\begin{equation}
\label{vuoto-attr}
{\cal Q}_{+}(-\infty )=\pm 2(\pi -\gamma )\, .
\end{equation}
For the plateau value of \( Z \) we obtain
\[
Z_{+}(-\infty )=\pm (\pi -2\gamma )\, .\]
The value is not determined uniquely due to the contribution of wide roots,
but it can be fixed using information from the repulsive regime.

In the repulsive regime, the real root at the origin is a special one with \( Z'(0)<0 \).
The sources are composed of two normal holes and the special root at the origin,
all of which has Bethe quantum number zero. One of the holes moves to the left
and the other one to the right and so \( S^{\pm }=\frac{1}{2} \), while the special root is central.
This allows for a unique solution of the plateau equations:
\[
{\cal Q}_{+}(-\infty )=2\gamma \: ,\, Z_{+}(-\infty )=-\pi +2\gamma \: .\]
Let us suppose that the counting function \( Z \) is analytic as a function of \( \gamma  \) which
is plausible because examining the NLIE for this state we find that all terms
are analytic in the coupling. This determines which branch to choose for the
plateau values in the attractive regime. We obtain the final result (for both
the attractive and the repulsive regime)
\[
\Delta ^{\pm }=\frac{1}{4}\left( 1-\frac{\gamma }{\pi }\right) =\frac{1}{8R^{2}}=\Delta ^{\pm }_{\pm 1/2,0}\: ,\]
which are the conformal weights of the vertex operators \( V_{(\pm 1/2,0)} \). The actual UV limit
can be a linear combination of these operators; in any case, it is not contained
in the UV spectrum of the sG/mTh theory.

Let us analyse the solution briefly to see the mechanism responsible for the
appearance of the special sources in the repulsive regime. Since the source
term is zero in the attractive regime, we expect that the \( Z \) function is odd
in \( \vartheta  \) with the behaviour depicted in figure \ref{specials_vacuum} a). Passing through the free-fermion
point \( \gamma =\frac{\pi }{2} \) to the repulsive regime, we observe that the plateau values change sign
and the derivative of the function at the origin becomes negative (part b) of
figure \ref{specials_vacuum}) so the root located there turns into a special one. At the same time,
however, there appear two new holes, one left and the other right moving, with
Bethe quantum number \( 0 \). This phenomenon can be thought of as the splitting of
the root at the origin into a special root and two holes and is one of the general
mechanisms in which the special sources are generated. Another mechanism will
be explored when we examine the UV limit of the soliton-antisoliton states. 

\vspace{0.3cm}
{\begin{figure}
{\centering \begin{tabular}{ccc}
\includegraphics{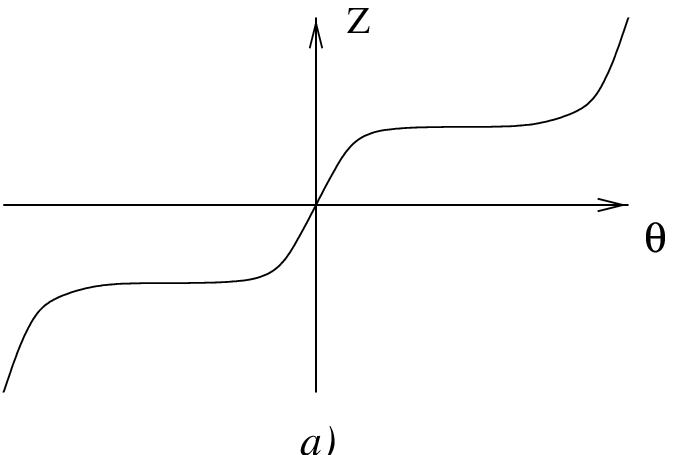} &
&
\includegraphics{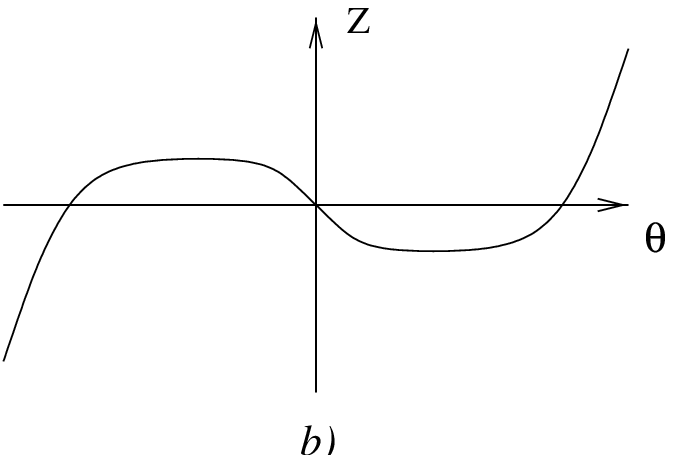} \\
\end{tabular}\par}

\caption{\small \label{specials_vacuum}The UV behaviour of \protect\( Z\protect \) for a) \protect\( \gamma >\frac{\pi }{2}\protect \) and b) \protect\( \gamma <\frac{\pi }{2}\protect \). In b), the root at the origin turned
into a special one and two normal holes appeared at the points where \protect\( Z\protect \) crosses
the horizontal axis with a positive derivative.}
\end{figure}\par}
\vspace{0.3cm}

\subsubsection{States containing two solitons}

As we have explained, for a generic configuration there are two possibilities,
corresponding to the half number of sites being even and odd. For the case of
two solitons (i.e. \( N_{H,eff}=N_{H}-2N_{S}=2 \)), the simplest possibilities are: 

\begin{enumerate}
\item Two holes quantised with \( I_{1,2}=\pm \frac{1}{2} \), \( \delta =0 \). Using the results of the previous subsection
yields: 
\[
\Delta ^{\pm }=\frac{1}{4(1-\gamma /\pi )}=\frac{R^{2}}{2}=\Delta ^{\pm }_{0,2}\: ,\]
corresponding to the operator \( V_{(0,2)} \), which is the UV limit of the lowest-lying two-soliton
state as it can also be seen from the TCS data. If instead of the minimal choice
\( I_{1,2} \) we take a nonminimal one \( I_{1}=I_{+} \), \( I_{2}=-I_{-} \) with \( I_{\pm }\in {\mathbb Z}_{+}-\frac{1}{2} \), we obtain
\[
\Delta ^{\pm }=\Delta _{0,2}^{\pm }+I_{\pm }-\frac{1}{2}\: ,\]
which corresponds to descendents of \( V_{(0,2)} \). This is a general phenomenon: the ``minimal''
choice of quantum numbers yields the primary state, while the nonminimal choices
give rise to descendents.
\item Two holes quantized with \( I_{1,2}=\pm 1 \), \( \delta =1 \), as proposed in \cite{ddv-97}. In the repulsive regime a special
root is required, as in the case of the \( \delta =1 \) vacuum state. The result (in both
regimes) is:
\[
\Delta ^{\pm }=1+\frac{1}{4}\frac{(\gamma /\pi )^{2}}{1-\gamma /\pi }=\Delta ^{+}_{1/2,2}=\Delta ^{-}_{1/2,2}+1\, .\]
This state is a linear combination of \( \bar{a}_{-1}V_{(1/2,2)} \) and \( a_{-1}V_{(-1/2,2)} \) which means that it is \emph{not}
contained in the local operator algebras of sG/mTh theories. 
\item Two holes quantized with \( I_{1}=0,\quad I_{2}=\pm 1 \), \( \delta =1 \). Consider in detail the case of \( I_{2}=+1 \) since the other
one is similar. Suppose that the hole with \( I_{1}=0 \) is a left mover and the other one
is a right mover (the other possibilities lead to a contradiction). We find
a solution to the plateau equation only in the attractive regime:
\[
{\cal Q}_{+}(-\infty )=\pm 2(\pi -\gamma ),\quad \: {\cal Q}_{-}(+\infty )=\pm 2(\pi -\gamma )\: .\]
In the repulsive regime the hole with quantum number \( 0 \) becomes a special hole
\( y \) and emits other two ordinary holes each quantised with zero. We obtain for
the plateau values
\[
{\cal Q}_{+}(-\infty )={\cal Q}_{-}(+\infty )=2\gamma \: .\]
The conformal weights turn out to be:
\[
\begin{array}{rl}
\displaystyle \Delta ^{+}= & 1+\displaystyle\frac{1}{4}\displaystyle\frac{(\gamma /\pi )^{2}}{1-\gamma /\pi }=\displaystyle\frac{1}{2}\left( \displaystyle\frac{1}{2R}+R\right) ^{2}\, ,\\
\Delta ^{-}= & \displaystyle\frac{1}{4}\displaystyle\frac{(\gamma /\pi )^{2}}{1-\gamma /\pi }=\displaystyle\frac{1}{2}\left( \displaystyle\frac{1}{2R}-R\right) ^{2}\, .
\end{array}\]
These are the conformal weights of the vertex operator \( V_{(1/2,2)} \). Performing a similar
calculation for \( I_{2}=-1 \) we obtain the weights of \( V_{(-1/2,2)} \). 
\end{enumerate}

\subsubsection{Generic number of solitons}

As the calculation proceeds in complete analogy with the two-soliton case, we
only give a summary of the results. The conformal families which we obtain depend
on how many holes move to the left and to the right. The primaries are obtained
for the minimal choice of the hole quantum numbers; increasing the quantum numbers
we obtain secondary states, as it was pointed out in the case with two holes.
The states we obtain are in the conformal family of a vertex operator \( V_{n,m} \) with
\[
m=N_{H,eff}\: ,\: n\in {\mathbb Z}+\frac{\delta }{2}\: .\]
As a consequence, all states with \( \delta =0 \) are contained in the UV spectrum of the
sG/mTh theory while the ones with \( \delta =1 \) are not, in agreement with the relation
(\ref{nmod2}). The complete expression for \( n \) is somewhat complicated, but we give it in
the case of symmetric ( \( N^{+}_{H}=N^{-}_{H}=N_{H,eff}/2 \) ) configurations:
\[
\begin{array}{ll}
n=0\: , & \delta =0\: ,\\
n=\pm \displaystyle\frac{1}{2}\: , & \delta =1\: .
\end{array}\]

\subsection{Examples with complex roots}

\subsubsection{\label{example_2handcp}Two holes and a close pair}

We now turn ourselves to the state with two holes and a close pair (in the repulsive
regime), which describes the antisymmetric soliton-antisoliton two-particle
state. We recall that according to the results from the IR asymptotics, the
quantum numbers of the complex roots can be \( 0 \) or \( \pm \frac{1}{2} \). Therefore, we are left two
possibilities: either the holes are quantised with integers or with half-integers. 

Let us start with the half-integer choice \( \delta =0 \) and suppose that one of the holes
is right moving (which is the case if its quantum number \( I_{1}=I_{+} \) is positive) and
the other one is left moving (i.e. \( I_{2}=I_{-}<0 \)). Using the general formula (\ref{plateau_solution}) and \( S^{\pm }=\frac{1}{2} \),
\( S=0 \), by a simple calculation we obtain
\[
{\cal Q}_{+}(-\infty )=Z_{+}(-\infty )=-2(\pi -2\gamma )\: ,\]
which is valid for \( \frac{\pi }{4}<\gamma <\frac{\pi }{2} \) (the complex pair remains central). The other plateau values
follow by oddity of the function \( Z \). For the other half of the repulsive regime
we have to include two new normal holes, one moving to the left and the other
to the right, and two special holes that remain central (a justification for
this will be given shortly) so that \( S^{\pm }=1 \). We then find 
\[
{\cal Q}_{+}(-\infty )=4\gamma \: ,\: Z_{+}(-\infty )=-2(\pi -2\gamma )\: ,\]
which is valid for \( \gamma \leq \frac{\pi }{4} \). In both cases the result for the UV weights turns out
to be 
\[
\Delta ^{\pm }=1-\frac{\gamma }{\pi }\pm I_{\pm }-\frac{1}{2}=\frac{1}{2R^{2}}\pm I_{\pm }-\frac{1}{2}=\Delta _{(\pm 1,0)}\pm I_{\pm }-\frac{1}{2}\, .\]
The primary state is obtained for the minimal choice \( I_{+}=-I_{-}=\frac{1}{2} \) and coincides with a
linear combination of the vertex operators \( V_{(\pm 1,0)} \) which is correct for the state
to be included in the spectrum of the sG/mTh theory and agrees with the behaviour
of the \( (s\bar{s})_{_{-}} \) state observed from TCS. If both of the holes move to the right or
to the left, we obtain descendents of the identity operator, i.e. states in
the vacuum module of the UV CFT.

\begin{figure}
{\centering \begin{tabular}{ccc}
\includegraphics{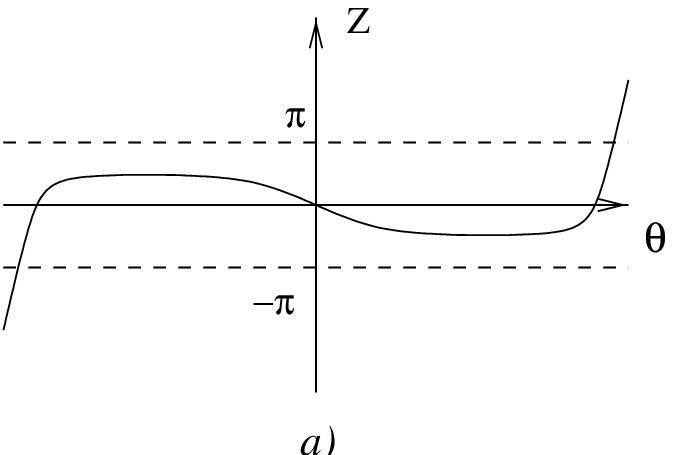} &
&
\includegraphics{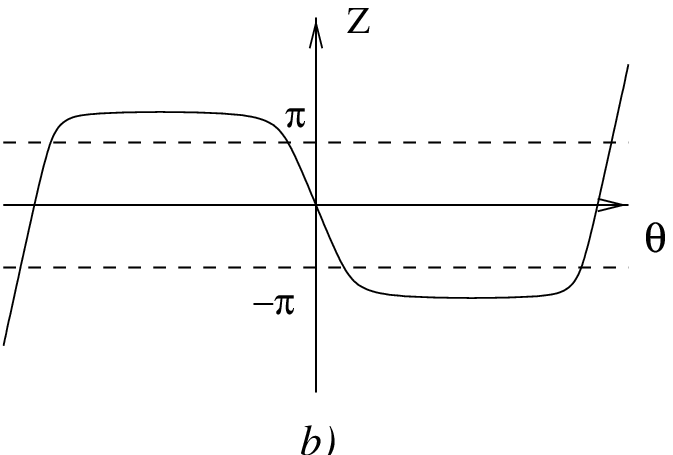} \\
\end{tabular}\par}

\caption{\small \label{twoholescp_specials}The UV behaviour of \protect\( Z\protect \) for a) \protect\( \gamma >\frac{\pi }{4}\protect \) and b) \protect\( \gamma <\frac{\pi }{4}\protect \).}
\end{figure}

Figure \ref{twoholescp_specials} shows how the special holes are generated in this case. Starting from
\( \gamma >\frac{\pi }{4} \) and decreasing the value of \( \gamma  \) the UV asymptotic form of the counting function
varies analytically. Since the real roots/holes are quantised by half-integers,
they are at positions where the function \( Z \) crosses a value of an odd multiple
of \( \pi  \). As the plots demonstrate, the behaviour of \( Z \) is in fact continuous at
the boundary \( \gamma =\frac{\pi }{4} \): it is our interpretation in terms of the sources that changes,
exactly because we try to keep the logarithm in the integral term of the NLIE
in its fundamental branch. The price we pay is that we have to introduce two
new normal holes (one moves left and the other one moves right) and two special
holes which are central.

Let us comment on the extension to the attractive regime. The complex root configuration
changes only by the possible presence of wide pairs (see subsection \ref{IRlimit_neutral}), which
however have no effect on the plateau values (\ref{plateau_solution}) as their contribution can be
absorbed in redefinition of the values \( k_{\pm } \). The other effect the wide roots have
is to shift the terms \( \Sigma _{\pm } \) by multiples of \( 2\pi  \), but this affects only the integers
\( N_{\pm } \) in (\ref{conformal}). Therefore we again obtain states which are descendents of \( V_{(\pm 1,0)} \).

For the state with integer quantisation of the holes, we only give the result.
If one of the holes is moving left and the other one to the right, we obtain
\[
\Delta ^{\pm }=\frac{1}{4}\left( 1-\frac{\gamma }{\pi }\right) \pm I_{\pm }=\frac{1}{8R^{2}}\pm I_{\pm }=\Delta _{(\pm 1/2,0)}\pm I_{\pm },\]
i.e. some linear combinations of descendents of the vertex operators \( V_{(\pm 1/2,0)} \). If both
of the holes move to the left or to the right, we obtain other descendants of
the same primary states. Again, this excludes the integer quantised states from
the spectrum of sG/mTh theory.

\subsubsection{Two holes and a selfconjugate complex root}

Taking the integer quantised state, the plateau values are given by the same
formulas as for the case with the close pair above, since the plateau equation
is identical with \( \delta _{eff}=\delta +1\bmod 2=0 \). The only difference is that the numbers \( q^{\pm }_{wide} \) in (\ref{sigmapm}) take a
nonzero value

\[
q^{\pm }_{wide}=-S^{\pm }.\]
If the one of the holes is a right mover, while the other one is a left mover,
the conformal weights turn out to be
\[
\Delta ^{\pm }=1-\frac{\gamma }{\pi }\pm I_{\pm }-1=\Delta _{(\pm 1,0)}\pm I_{\pm }-1.\]
If the two holes move in the same direction, we again obtain secondaries of
the vacuum state. 

Concerning the extension to the attractive regime, one can again check that
the plateau solution remains unchanged for the root configuration described
in subsection \ref{IRlimit_neutral}, therefore the conformal family to which the state belongs remains
the same, similarly to the case of the \( (s\bar{s})_{-} \) state.

One can ask what happens if we quantize with half-integers. The result is, similarly
to the case with the complex pair, that we obtain descendants of the operators
\( V_{(\pm 1/2,0)} \). Therefore we conclude that in this case the integer quantised configuration
must be accepted, while the half-integer one is ruled out, in full accordance
with the rule \( \delta _{eff}=0 \).

\subsubsection{Four hole examples}

We do not discuss all the possible examples of four holes with complex roots.
Instead, we concentrate only on two cases that will come up in the comparison
we make with TCS data in section \ref{NLIE_and_TCS}.

The first example is four holes with a complex pair. We choose the half-integer
quantised prescription and the hole quantum numbers \( I_{1}=-I_{2}=\frac{1}{2} \) and \( I_{3}=-I_{4}=\frac{3}{2} \), which is the minimal
choice with two left moving and two right moving holes (the nonminimal choices
lead again to descendents). Performing the calculation we obtain that
\begin{equation}
\label{4holescp_delta}
\Delta ^{\pm }=\displaystyle\frac{(\pi -2\gamma )^{2}}{4\pi (\pi -\gamma )}+2=\displaystyle\frac{R^{2}}{2}+\displaystyle\frac{1}{2R^{2}}+1\, .
\end{equation}
The operators having these conformal weights are the following:
\[
\bar{a}_{-2}V_{(2,1)}\: ,\bar{a}_{-1}\bar{a}_{-1}V_{(2,1)}\: ,a_{-2}V_{(2,-1)}\: \mathrm{and}\: a_{-1}a_{-1}V_{(2,-1)}\: .\]
The UV limit of the scaling functions corresponds to some linear combination
of these states. This is also confirmed by comparison with TCS data. The integer
quantisation leads again to states not present in the UV spectrum. Considering
four holes with a selfconjugate root in the middle, we arrive at similar result,
with the difference that this time the integer quantisation leads to the correct
weights (\ref{4holescp_delta}), while the half-integer prescription yields states which are not
part of sG/mTh Hilbert space.

\section{\label{on_TCS}Truncated Conformal Space at \protect\( c=1\protect \)}

\setcounter{equation}{0}

Before presenting the numerical results, let us give a brief description of
Truncated Conformal Space (TCS) method for \( c=1 \) theories. The TCS method was originally
created to describe perturbations of Virasoro minimal models in finite spatial
volume \cite{yurzam}. Here we present an extension of the method to study perturbations
of a \( c=1 \) compactified boson, more closely the perturbation corresponding to sine-Gordon
theory.

The Hilbert space of the \( c=1 \) theory can be split into sectors labelled by the
values of \( P \) and \( Q \), which are quantised by integers. In the numerical computations
we shall restrict ourselves to the \( P=0 \) sector (it is not expected that any relevant
new information would come from considering \( P\neq 0 \)). The TCS method consists of retaining
only those states in such a sector for which the eigenvalue of \( H_{CFT} \) is less than
a certain upper value \( E_{cut} \), so the truncated space is defined as

\begin{equation}
\label{}
{\cal H}_{TCS}(s,m,E_{cut})=\left\{ \left| \Psi \right\rangle :\: P\left| \Psi \right\rangle =s\left| \Psi \right\rangle ,Q\left| \Psi \right\rangle =m\left| \Psi \right\rangle ,H_{CFT}\left| \Psi \right\rangle \leq E_{cut}\left| \Psi \right\rangle \right\} \: .
\end{equation}

For a given value of \( s \), \( m \) and \( E_{cut} \) this space is always finite dimensional. In
this space, the Hamiltonian is represented by a finite size matrix whose eigenvalues
can be computed using a numerical diagonalization method. The explicit form
of this matrix is the following:

\begin{equation}
\label{}
\widehat{H}=\displaystyle\frac{2\pi }{L}\left( \widehat{L_{0}}+\widehat{\overline{L}_{0}}-\displaystyle\frac{c}{12}\widehat{I}+\lambda \displaystyle\frac{L^{2-h}}{\left( 2\pi \right) ^{1-h}}\widehat{B}\right) \: ,
\end{equation}
where \( \widehat{L_{0}} \) and \( \widehat{\overline{L_{0}}} \) are diagonal matrices with their diagonal elements being the left
and right conformal weights, \( \widehat{I} \) is the identity matrix,

\begin{equation}
\label{}
h=\Delta ^{+}_{V}+\Delta ^{-}_{V}=\displaystyle\frac{\beta ^{2}}{4\pi }=\displaystyle\frac{2p}{p+1}
\end{equation}
is the scaling dimension of the perturbing potential \( V \) and the matrix elements
of \( \widehat{B} \) are

\begin{equation}
\label{}
\widehat{B}_{\Phi ,\Psi }=\displaystyle\frac{1}{2}\left\langle \Phi \right| V_{(1,0)}(1,1)+V_{(0,1)}(1,1)\left| \Psi \right\rangle \: .
\end{equation}
We choose our units in terms of the soliton mass \( {\cal M} \) which is related to the coupling
constant \( \lambda  \) by the mass gap formula obtained from TBA in \cite{mass_scale}:

\begin{equation}
\label{mass_gap}
\lambda =\kappa {\cal M}^{2-h},
\end{equation}
where

\begin{equation}
\label{}
\kappa =\displaystyle\frac{2\Gamma (h/2)}{\pi \Gamma (1-h/2)}\left( \displaystyle\frac{\sqrt{\pi }}{2\Gamma \left( \displaystyle\frac{1}{2-h}\right) \Gamma \left( \displaystyle\frac{h}{4-2h}\right) }\right) ^{2-h}\, .
\end{equation}
In what follows we normalize the energy scale by taking \( {\cal M}=1 \) and denote the dimensionless
volume \( {\cal M}L \) by \( l \). For numerical computations, we shall use the dimensionless Hamiltonian

\begin{equation}
\label{dimlessham}
\widehat{h}=\displaystyle\frac{\widehat{H}}{{\cal M}}=\displaystyle\frac{2\pi }{l}\left( \widehat{L_{0}}+\widehat{\overline{L}_{0}}-\displaystyle\frac{c}{12}\widehat{I}+\kappa \displaystyle\frac{l^{2-h}}{\left( 2\pi \right) ^{1-h}}\widehat{B}\right) \: .
\end{equation}

The usefulness of the TCS method lies in the fact that it provides a nonperturbative
method of numerically obtaining the spectrum (the mass gap, the mass ratios
and the scattering amplitudes) of the theory. Therefore it can serve as a tool
to check the exact results obtained for integrable field theories and get a
picture of the physical behaviour even for the nonintegrable case. The systematic
error introduced by the truncation procedure is called the \emph{truncation
error}, which increases with the volume \( L \) and can be made smaller by increasing
the truncation level (at the price of increasing the size of the matrices, which
is bound from above by machine memory and computation time).

Let us make some general remarks on how the TCS method applies to \( c=1 \) theories.
First note that the Hilbert space (even after specifying the sector by the eigenvalues
of \( P \) and \( Q \)) consists of infinitely many Verma modules labelled by the quantum
number \( n \). At any finite value of \( E_{cut} \) only finitely many of such Verma modules
contribute, but their number increases with \( E_{cut} \). As a result one has to deal with
many more states than in the case of minimal models. The results of TCS are
supposed to approach the exact results in the limit \( E_{cut}\: \rightarrow \: \infty  \), and the convergence can
be very slow, while the number of states rises faster than exponentially with
the truncation level. The perturbing operator has scaling dimension which ranges
between \( 0 \) and \( 2 \), becoming more relevant in the attractive regime, while the
number of states corresponding to a given value of \( E_{cut} \) becomes larger as moving
towards \( p=0 \), which affects the convergence just the other way around. 

Generally, the energy of any state goes with the volume \( L \) as

\begin{equation}
\label{scalingfun}
\displaystyle\frac{E_{\Psi }(L)}{M}=-\displaystyle\frac{\pi \left( c-12\left( \Delta _{\Psi }+\overline{\Delta }_{\Psi }\right) \right) }{6l}+Bl+\displaystyle\sum ^{\infty }_{k=1}C_{k}\left( \Psi \right) l^{k(2-h)}\: ,
\end{equation}
where \( \Delta _{\Psi } \) (\( \overline{\Delta }_{\Psi } \)) are the left (right) conformal dimensions of the state in the ultraviolet
limit, \( B \) is the universal bulk energy constant (the vacuum energy density) and
the infinite sum represents the perturbative contributions from the potential
\( V \). 

The bulk energy constant has also been predicted from TBA and reads \cite{mass_scale}

\begin{equation}
\label{bulk}
B=-\displaystyle\frac{1}{4}\tan \left( \displaystyle\frac{p\pi }{2}\right) 
\end{equation}
(the same result was obtained from the NLIE approach in \cite{ddv-95}). This is a highly
nonanalytic function of \( p \) and it becomes infinite at the points where \( p \) is an
odd integer. In fact, at these points there is a value of \( k \) for which \( k(2-h)=1 \), and
\( C_{k}\left( \Psi \right) \: \rightarrow \: \infty  \). The infinite parts of \( B \) and \( C_{k}\left( \Psi \right)  \) exactly cancel, leaving a logarithmic (proportional
to \( l\: \log l \)) and a finite linear contribution to the energy, by a sort of a resonance
mechanism. All of these ``logarithmic points'' are in the repulsive regime. However,
due to UV problems in the repulsive regime we are not able to check numerically
the logarithmic corrections to the bulk energy.

The origin of UV divergences can be understood from \emph{conformal perturbation
theory} (CPT). It is known that when the scaling dimension \( h \) of the perturbing
potential exceeds \( 1 \), CPT suffers from ultraviolet divergences which should be
removed by some renormalization procedure. The TCS method is something very
similar to CPT: it operates in the basis of the UV wave functions as well, but
computes the energy levels using the variational approach and therefore could
be called ``\emph{conformal variation theory}'' \emph{}(CVT). As a result, we
expect that there could be UV divergences for the range of couplings where \( h>1 \)
which is exactly the repulsive regime \( p>1 \) \cite{kl-me2}. The numerical analysis has in fact
shown that in the repulsive regime the TCS energy eigenvalues did not converge
at all when increasing the truncation level. 

Fortunately, there exists a way out: since we expect to find a sensible quantum
field theory when the UV cutoff is removed, it should be the case that the \emph{relative
energy levels} \( {\cal E}_{\Psi }(L)=E_{\Psi }(L)-E_{vac}(L) \) converge to some limit. This is exactly the behaviour that
we observed. Consequently, in the repulsive regime one can only trust the \textit{relative}
scaling functions produced by the TCS method, while in the attractive regime
even the \textit{absolute} energy values agree very well with the predictions
of the NLIE \cite{noi3} (including the predicted bulk energy constant (\ref{bulk}), which is completely
analytic for \( p<1 \) and thus logarithmic corrections are absent as well). 

Our numerical results show that the smaller the value of \( p \) is the faster the
convergence is (with the understanding that in the repulsive regime by convergence
of TCS we mean the convergence of the energies relative to the vacuum). On the
other hand, even in the attractive regime the convergence is so slow that to
get reliable results (which means errors of order \( 10^{-3}-10^{-2} \) for the volume \( l \) ranging
from \( 0 \) to somewhere between \( 5 \) and \( 10 \)) one has to work with dimensions around
\( 4000 \). This means that the TCS for \( c=1 \) theories is far less convergent than the one
for minimal models (in the original Lee-Yang example the authors of \cite{yurzam} took a
\( 17 \) dimensional Hilbert space (!) and arrived to very accurate results).

\section{\label{NLIE_and_TCS} Comparison of the NLIE to TCS}

\setcounter{equation}{0}

In this section we give the numerical comparison between the NLIE solutions
and the TCS data. In the IR limit of the NLIE gives back the scattering theory
of the sG/mTh model. In this limit, the scaling functions are described by the
asymptotic BA, which for two-particle states can be written as follows

\begin{eqnarray}
l\sinh \vartheta _{1}-i\log S(\vartheta _{1}-\vartheta _{2})=2\pi I_{1}, &  & \nonumber \\
l\sinh \vartheta _{2}-i\log S(\vartheta _{2}-\vartheta _{1})=2\pi I_{2}, &  & \label{IRBA} 
\end{eqnarray}
where \( \vartheta _{1,2} \) are the rapidities of the two particles and the energy of the state
relative to the vacuum is given by

\begin{equation}
\label{}
E_{I_{1}I_{2}}(l)=m_{1}\cosh \vartheta _{1}+m_{2}\cosh \vartheta _{2},
\end{equation}
with the understanding that one has to substitute for \( \vartheta _{1,2} \) the solutions of the
equations (\ref{IRBA}). The corrections from the NLIE to the IR asymptotic Bethe Ansatz
of (\ref{IRBA}) decrease exponentially with \( l \). This is also expected on physical basis:
the theory has a finite mass gap and hence these finite size corrections should
decay exponentially with the volume. Therefore to see a substantial effect of
the NLIE as compared to the asymptotic BA one has to go down to small values
of the volume. This is also advisable for the comparison with TCS, because the
truncation effects are smaller if one is closer to the UV region. 

On the other hand, the UV dimensions computed by the ``kink approximation'' were
shown to be correct. Therefore the comparison between TCS and NLIE serves in
fact for verifying that the scaling functions obtained from the NLIE interpolate
correctly between the UV and IR limits in the intermediate range of scales.
Since the accuracy of TCS is getting worse for larger volumes and, on the other
hand, the iterative solution of the NLIE becomes problematic for small volumes,
it is at intermediate scales that a useful comparison can be made. 

To have a comparison between TCS and the NLIE, we have to get numerical predictions
for the scaling functions from the NLIE. Before presenting the results we briefly
discuss the method used for the numerical solution of NLIE.

We recall that the NLIE (\ref{nlie-cont}) contains a source term \( g(\vartheta |\vartheta _{j}) \) which is specified by giving
the root/hole structure of the given state and depends on the positions of the
holes, special roots/holes and complex roots. These positions in turn are fixed
by the Bethe quantisation conditions. The coupled nonlinear equations coming
from the integral equation supplemented with the quantisation conditions can
be solved numerically by an iterative procedure. First one chooses a starting
position for the positions of the sources \( \vartheta _{j} \), which we set to the asymptotic
position for the complex roots and to zero for all hole positions. Then one
iterates the integral equation (using fast Fourier transform to evaluate the
convolution) to update the counting function \( Z \). Using this new \( Z \), an improved
determination of position of the \( \vartheta _{j} \) can be obtained, and is fed back into the
integral equation for a new iteration cycle. The process is repeated until the
solution is found to a prescribed precision ( usually \( 10^{-6} \) ). For large \( l \), the
source term dominates, while the correction coming from the integral term is
exponentially small. Therefore we expect that the further we go to the IR regime
the faster the iteration converges which does in fact hold. Hence it is preferable
to start iterating at the largest desired value of \( l \) and decrease the volume
gradually, always taking as a starting point at the next value of the volume
the solution found at the previous value.

We do not present results for pure hole states which have already been discussed
in a previous paper \cite{noi3}, both in the attractive and the repulsive regime. Instead
we concentrate on the new results obtained for states containing complex roots
and in the repulsive regime. 

Let us start with the \( (s\bar{s})_{\pm } \) states in table \ref{ssplusnum} and \ref{ssminusnum}, respectively. The NLIE results
can be considered to be exact to the accuracy showed, therefore the total deviation
must be attributed to truncation effects. Hence we expect that it shows a growing
tendency with the volume \( l \), which is in fact the case (for the lowest lying
\( (s\bar{s})_{\pm } \) states the NLIE and TCS predictions intersect each other around \( l=7.5 \) and \( l=6.0 \), respectively,
but this is a numerical artifact due to truncation errors from TCS). The order
of the deviations is consistent with what we expect from measuring the accuracy
of the mass gap from TCS. It also gets larger if we consider a state of higher
energy which is expected since we get closer to the truncation level. 

\begin{table}
{\centering \begin{tabular}{|c||c|c|c||c|c|c|}
\hline 
&
\multicolumn{3}{|c||}{ \( I_{1}=-I_{2}=1 \)}&
\multicolumn{3}{|c|}{ \( I_{1}=-I_{2}=2 \)}\\
\hline 
\hline 
\( l \)&
TCS&
NLIE&
Dev. (\%)&
TCS&
NLIE&
Dev. (\%)\\
\hline 
0.5&
15.4665&
 15.4152&
0.33&
40.3852&
 40.2870&
0.24\\
\hline 
1.0&
8.0919&
8.0485&
0.54&
 20.3794 &
 20.2325&
0.73\\
\hline 
1.5&
5.7245&
5.6729&
0.9&
13.7695&
13.5837&
1.4\\
\hline 
2.0&
4.5887&
4.5325&
1.2&
10.5064&
10.2886&
2.1\\
\hline 
2.5&
3.9358&
3.8782&
1.5&
8.5782&
8.3347&
2.9\\
\hline 
3.0&
3.5186&
3.4617&
1.6&
 7.3148&
7.0501&
3.8\\
\hline 
3.5&
3.2317&
3.1775&
1.7&
 6.4283&
6.1475&
4.6\\
\hline 
4.0&
3.0236&
2.9734&
1.7&
5.7750&
5.4820&
5.3\\
\hline 
4.5&
2.8662&
2.8211&
1.6&
5.2757&
4.9737&
6.1\\
\hline 
5.0&
2.7430&
2.7038&
1.4&
4.8826&
4.5749&
6.7\\
\hline 
5.5&
2.6437&
2.6113&
1.2&
4.5658&
4.2546&
7.3\\
\hline 
6.0&
2.5619&
2.5367&
1.0&
4.3054&
3.9931&
7.8\\
\hline 
6.5&
2.4932&
2.4756&
0.7&
 4.0878&
3.7762 &
8.3\\
\hline 
7.0&
2.4344&
2.4248&
0.4&
3.9034&
3.5941&
8.6\\
\hline 
7.5&
2.3834&
2.3820&
0.06&
3.7451&
3.4394&
8.9\\
\hline 
8.0&
2.3385&
2.3456&
0.3&
3.6077&
3.3068&
9.1\\
\hline 
8.5&
 2.2987&
2.3143&
0.7&
3.4874&
3.1921&
9.3\\
\hline 
9.0&
2.2629&
2.2872&
1.1&
 3.3811&
3.0923&
9.3\\
\hline 
9.5&
 2.2305&
2.2635&
1.5&
 3.2864&
3.0047&
9.4\\
\hline 
10.0&
2.2010&
2.2428&
1.8&
3.2017&
2.9275&
9.4\\
\hline 
\end{tabular}\par}

\caption{\small \label{ssplusnum} Numerical comparison of the NLIE prediction and the TCS data for the first
two \protect\( (s\overline{s})_{+}\protect \) states, at \protect\( p=1.5\protect \) .}
\end{table} 

\begin{table}
{\centering \begin{tabular}{|c||c|c|c||c|c|c|}
\hline 
&
\multicolumn{3}{|c||}{ \( I_{1}=-I_{2}=\displaystyle\frac{1}{2} \)}&
\multicolumn{3}{|c|}{\( I_{1}=-I_{2}=\displaystyle\frac{3}{2} \) }\\
\hline 
\hline 
\( l \)&
TCS&
NLIE&
Dev. (\%)&
TCS&
NLIE&
Dev. (\%)\\
\hline 
0.5&
15.0922&
15.0605&
0.21&
40.2909&
40.1949&
0.24\\
\hline 
1.0&
7.6401&
7.5916&
0.64&
20.2522&
20.1042&
0.73\\
\hline 
1.5&
5.2475&
5.1891&
1.1&
13.6247&
13.4355&
1.4\\
\hline 
2.0&
4.1198&
4.0565&
1.6&
10.3511&
10.1282&
2.2\\
\hline 
2.5&
3.4900&
3.4264&
1.8&
8.4171&
8.1667&
3.1\\
\hline 
3.0&
3.1009&
3.0408&
1.9&
7.1500&
6.8771&
4.0\\
\hline 
3.5&
2.8428&
2.7889&
1.9&
6.2620&
5.9710&
4.9\\
\hline 
4.0&
2.6621&
2.6163&
1.8&
5.6076&
5.3039&
5.7\\
\hline 
4.5&
2.5298&
2.4934&
1.5&
5.1082&
4.7952&
6.5\\
\hline 
5.0&
2.4292&
2.4032&
1.1&
4.7158&
4.3967&
7.3\\
\hline 
5.5&
2.3502&
2.3352&
0.65&
4.4000&
4.0779&
7.9\\
\hline 
6.0&
2.2864&
2.2827&
0.17&
4.1410&
3.8184&
8.5\\
\hline 
6.5&
2.2337&
2.2415&
0.35&
3.9250&
3.6041&
8.9\\
\hline 
7.0&
2.1892&
2.2085&
0.88&
3.7422&
3.4251&
9.3\\
\hline 
7.5&
2.1509&
2.1818&
1.4&
3.5858&
3.2738&
9.5\\
\hline 
8.0&
2.1175&
2.1599&
2.0&
3.4503&
3.1450&
9.7\\
\hline 
8.5&
2.0879&
2.1417&
2.5&
3.3318&
3.0343&
9.8\\
\hline 
9.0&
2.0614&
2.1264&
3.1&
3.2274&
2.9385&
9.8\\
\hline 
9.5&
2.0374&
2.1134&
3.6&
3.1347&
2.8551&
9.8\\
\hline 
10.0&
2.0156&
2.1023&
4.1&
3.0517&
2.7821&
9.7\\
\hline 
\end{tabular}\par}

\caption{\small \label{ssminusnum} Numerical comparison of the NLIE prediction and the TCS data for the first
two \protect\( (s\overline{s})_{-}\protect \) states, at \protect\( p=1.5\protect \) .}
\end{table} 

It is also possible to consider four hole states with complex roots (see table
\ref{4holesnum}). In this case we consider the lowest example of four holes with a selfconjugate
wide root resp. with a close pair, with the prescription of integer resp. half-integer
quantisation. 

\begin{table}
{\centering \begin{tabular}{|c||c|c|c||c|c|c|}
\hline 
&
\multicolumn{3}{|c||}{4 holes \( (-2,-1,1,2) \) + s.c.w.}&
\multicolumn{3}{|c|}{4 holes \( \left( -\displaystyle\frac{3}{2},-\displaystyle\frac{1}{2},\displaystyle\frac{1}{2},\displaystyle\frac{3}{2}\right)  \)+ c.p.}\\
\hline 
\hline 
\( l \) &
TCS&
NLIE&
Dev. (\%)&
TCS&
NLIE&
Dev. (\%)\\
\hline 
0.5&
51.2829&
51.2224&
0.12&
50.7801&
50.7101&
0.14\\
\hline 
1.0&
26.2457&
26.1581&
0.33&
25.6191&
25.5111&
0.42\\
\hline 
1.5&
18.0383&
17.9318&
0.59&
17.3879&
17.2521&
0.79\\
\hline 
2.0&
14.0138&
13.8965&
0.83&
13.3812&
13.2257&
1.2\\
\hline 
2.5&
11.6458&
11.5258&
1.0&
11.0490&
10.8823&
1.5\\
\hline 
3.0&
10.0998&
9.9802&
1.2&
9.5416&
9.3709&
1.8\\
\hline 
3.5&
9.0162&
8.9016&
1.3&
8.4957&
8.3272&
2.0\\
\hline 
4.0&
8.2162&
8.1101&
1.3&
7.7318&
7.5706&
2.1\\
\hline 
4.5&
7.5924&
7.5086&
1.1&
7.1518&
7.0015&
2.1\\
\hline 
5.0&
7.1110&
7.0380&
1.0&
6.6972&
6.5609&
2.1\\
\hline 
\end{tabular}\par}

\caption{\small \label{4holesnum} Numerical comparison for the first four hole states containing a selfconjugate
wide root resp. a closed pair, with the four holes having the quantum numbers
indicated in the table. }
\end{table} 

The reason to confine ourselves to the lowest choices for the quantum numbers
is related to the problem of identifying the states, since the higher we go
in energy the denser the spectrum gets. Given that the results of TCS are approximate,
it becomes very hard to identify the points in the TCS plot which form a line
corresponding to a given energy level. This job can be done by using BA combined
with other information. For example, it can be observed that the states \( (s\bar{s})_{+} \) and
\( (s\bar{s})_{-} \) always come in pairs close to each other, the symmetric state being higher.
These energy levels cannot cross each other because the sG/mTh theory is parity
conserving and these states have opposite parity. This information is enough
to identify the first two such pairs up to \( l=10 \). The four-particle states in table
\ref{4holesnum} lie in the sector with topological charge \( Q=2 \), where the first few energy levels
are soliton-soliton two-particle states. Eliminating these states from the TCS
spectrum (using asymptotic BA or even full NLIE predictions) facilitates the
identification of the required lines. The reason why the data presented in table
\ref{4holesnum} only go up to \( l=5 \) is that after this point the TCS points corresponding to the
given state cannot be identified unambiguously. For the same reason we do not
present four hole states with higher quantum numbers. Although there are TCS
data points which agree with the NLIE predictions within the estimated truncation
errors, the density of the TCS points is so high that it makes impossible to
identify the correct ones.

\section{Conclusions}

\setcounter{equation}{0}

In this paper we have studied how the nonlinear integral equation deduced from
the light cone lattice model of \cite{ddv-87} describes the excited states of the sine-Gordon/massive
Thirring theory. We can summarize the new results as follows:

\begin{enumerate}
\item We have presented a new derivation of the fundamental NLIE from the light cone
lattice which avoids the problems originating from the multivaluedness of the
complex logarithm function and gives us a correct form of the NLIE. 
\item By examining the infrared limit of our equation we have shown that it leads
to the correct two-particle S-matrices of sine-Gordon theory.
\item By computing the UV conformal weights from the NLIE we have shown that it is
consistent with the UV spectrum of sG/mTh theory only if we choose the parameter
\( \delta  \) (i.e. the quantisation rule) in a way different from the proposal of \cite{ddv-97}. The
new rule is that \( \delta  \) should be equal to the number of the self-conjugate roots
modulo \( 2 \).
\item We have verified the predictions of the NLIE by comparing them to results coming
from the TCS approach.
\end{enumerate}
In view of our results we think that the philosophy behind the NLIE has to be
changed. Namely, the NLIE describes nothing else than a collection of scaling
functions including ones not realized in the quantum field theory we wish to
study. We have seen that the spectrum provided by the NLIE is much larger than
necessary for sG/mTh, and the correct approach is that from the variety of scaling
functions predicted by the NLIE one has to select the ones corresponding to
the Hilbert space of the quantum field theory to be described. The key condition
in performing the selection is the \emph{locality} of the resulting operator
algebra (which we established at the UV fixed point, using the results of \cite{kl-me}).
With this point understood, our results provide a strong evidence that the NLIE
\emph{does} in fact describe the finite volume spectrum of sG/mTh theory (for
even value of the topological charge).

Let us discuss briefly the relation between the TBA approach to excited states
\cite{tateo_dorey} and the NLIE approach. The TBA approach is based on an analytic continuation
of the vacuum TBA equation to the complex plane of the volume parameter \( l \). By
encircling certain branch points in that plane, it is possible to obtain the
equation describing the evolution of some excited state scaling function and
repeating the procedure other states can be found. At present  this is the most
powerful approach known to obtain excited state scaling functions in perturbed
minimal models. The main drawback of the method is that it is difficult to compute
the equation for a generic excited state.

The NLIE approach has the advantage that it can give the equation for all of
the excited states in closed form. The price we pay is that it is a complex
nonlinear integral equation while the TBA equations are real, therefore its
structure is much more complicated due to the fact that it encodes a lot of
information in compact form. Another limitation is that it requires the knowledge
of an integrable lattice regularisation. On the other hand, this has a positive
consequence too: we do not put in by hand the model to be described as in TBA,
but it emerges during the derivation (see the comments about the appearance
of the sine-Gordon phaseshift in section \ref{NLIE_derivation} after equation (\ref{chi})). 

Finally, let us point to some open questions:

\begin{enumerate}
\item Is the set of scaling functions provided by the NLIE complete i.e. can we find
to every sG/mTh state (of even topological charge) a solution of the NLIE describing
its finite volume behaviour? This can be called a ``counting problem''. The main
difficulty is that the structure of the solutions is highly dependent on the
value of the coupling constant -- to see that it is enough to consider e.g.
the appearance of special sources. There should also be dependence on the coupling
related to physical reasons (e.g. breather thresholds -- cf. subsection \ref{IRlimit_neutral}).
\item From the form of the source terms in the NLIE it seems likely that the excited
state equations can be obtained by an analytic continuation procedure analogous
to the one used in TBA \cite{tateo_dorey} to obtain the excited state TBA equations. Certain
features of the arrangement of the complex roots in the attractive regime and
their behaviour at breather thresholds also point into this direction. This
an interesting question to investigate because it can shed light on the organization
of the space of states and can lead closer to solving the counting problem described
above. 
\item It was conjectured in \cite{noi} that a so-called \( \alpha  \)-twisted version (\`{a} la Zamolodchikov
\cite{mass_scale}) of the NLIE can lead to excited state scaling functions of \( \Phi _{(1,3)} \) perturbations
of minimal models which could provide a link between the NLIE and the TBA approach
to excited states. In \cite{noi} it was found that the vacuum scaling functions obtained
in this way agreed with the ones predicted by TBA within a very small error.
Therefore studying the \( \alpha  \)-twisted NLIE is clearly an interesting problem.
\item The NLIE technique so far has been limited to states with even topological charge.
A work on the description of the odd sector and of the sine-Gordon - massive
Thirring difference is under way, the results of which will be published soon
\cite{preparation}.
\end{enumerate}
The second and third points are also important because their investigation may
lead closer to understanding the relation between the TBA and the NLIE approaches.
It is quite likely that establishing a connection between the two methods would
facilitate the development of both and may point to some common underlying structure.

\vspace{0.5cm}
\textbf{Acknowledgements -} We are indebted to P. E. Dorey, V. A. Fateev and
E. Quattrini and especially to C. Destri for useful discussions and comments.
This work was supported in part by NATO Grant CRG 950751, by European Union
TMR Network FMRX-CT96-0012 and by INFN \emph{Iniziativa Specifica} TO12. G.
T. has been partially supported by the FKFP 0125/1997 and OTKA T016251 Hungarian
funds.


\begin{thebibliography}{10}
\bibitem[1]{YY}C.N. Yang and C.P. Yang, \textit{J. Math. Phys.} \textbf{10} (1969) 1115.
\bibitem[2]{Zam-tba1}Al.B. Zamolodchikov, \textit{Nucl. Phys.} \textbf{B342} (1990) 695-720.
\bibitem[3]{Fendley}P. Fendley, \emph{Nucl. Phys.} \textbf{B374} (1992) 667-691, hep-th/9109021. 
\bibitem[4]{tateo_dorey}P. Dorey and R. Tateo, \emph{Nucl. Phys.} \textbf{B482} (1996) 639-659, hep-th/9607167.
\\
P. Dorey and R. Tateo, \emph{Nucl. Phys.} \textbf{B515} (1998) 575-623, hep-th/9706140. 
\bibitem[5]{blz}V.V. Bazhanov, S.L. Lukyanov and A.B. Zamolodchikov, \emph{Nucl. Phys.} \textbf{B489}
(1997) 487-531, hep-th/9607099. 
\bibitem[6]{susy}Paul Fendley, \emph{Adv. Theor. Math. Phys.} \textbf{1} (1998) 210-236, hep-th/9706161. 
\bibitem[7]{yurzam}V.P. Yurov and A.B. Zamolodchikov, \textit{Int.J.Mod.Phys.} \textbf{A5} (1990)
3221-3246. 
\bibitem[8]{ddv-87}C. Destri and H.J. De Vega, \emph{Nucl. Phys.} \textbf{B290} (1987) 363-391. 
\bibitem[9]{ddv-92}C. Destri and H.J. De Vega, \textit{Phys. Rev. Lett.} \textbf{69} (1992) 2313-2317. 
\bibitem[10]{ddv-95}C. Destri and H.J. De Vega, \textit{Nucl. Phys.} \textbf{B438} (1995) 413-454,
hep-th/9407117.
\bibitem[11]{klumper}\noindent A. Klümper and P.A. Pearce, \textit{J. Stat. Phys.} \textbf{64} (1991)
13; \\
A. Klümper, M. Batchelor and P.A. Pearce, \textit{J. Phys.} \textbf{A24} (1991)
3111.
\bibitem[12]{noi}D. Fioravanti, A. Mariottini, E. Quattrini and F. Ravanini, \textit{Phys. Lett.}
\textbf{B390} (1997) 243-251, hep-th/9608091.
\bibitem[13]{ddv-97}C. Destri and H. De Vega, \textit{Nucl. Phys.} \textbf{B504} (1997) 621-664,
hep-th/9701107.
\bibitem[14]{noi3}G. Feverati, F. Ravanini and G. Takács: \emph{Truncated Conformal Space at c=1,
Nonlinear Integral Equation and Quantisation Rules} \emph{for Multi-Soliton
States,} preprint DFUB-98-04, hep-th/9803104. To appear in \emph{Physics Letters}
\textbf{B}. 
\bibitem[15]{mariottini}A. Mariottini, \emph{Ansatz di Bethe Termodinamico ed Equazione di Destri-de
Vega in Teorie di Campo Bidimensionali} (in Italian), M. Sc. thesis -- University
of Bologna (March 1996), available from http://www-th.bo.infn.it/hepth/papers.html
.
\bibitem[16]{zinn-justin}P. Zinn-Justin, \emph{Nonlinear Integral Equations for Complex Affine Toda Models
Associated To Simply Laced Lie Algebras}, preprint LPTENS-97-65, hep-th/9712222. 
\bibitem[17]{kl-me}T. Klassen and E. Melzer, \emph{Int. J. Mod. Phys}\textbf{. A8} (1993) 4131-4174,
hep-th/9206114.
\bibitem[18]{ginsparg}P. Ginsparg, \emph{Nucl. Phys.} \textbf{B295} (1988) 153-170.  
\bibitem[19]{mass_scale}Al.B. Zamolodchikov, \textit{Int. J. Mod. Phys.} \textbf{A10} (1995) 1125-1150. 
\bibitem[20]{kl-me2}T.R. Klassen and E. Melzer, \emph{Nucl. Phys.} \textbf{B370} (1992) 511-550.
\bibitem[21]{preparation}G. Feverati, F. Ravanini and G. Tak{\'a}cs, in preparation.
\end{thebibliography}
\end{document}